%% file: main.tex
\documentclass{article}
\usepackage[T1]{fontenc}
\usepackage{textcomp}
\usepackage{multicol}
\usepackage{multirow}
\usepackage{array}
\usepackage{tabularx}
\usepackage{graphicx}
\usepackage{hyperref}
\usepackage[export]{adjustbox}

\usepackage{textcomp}
\usepackage{listings}
\usepackage{xcolor}

\usepackage{makecell}

\usepackage[normalem]{ulem}

\usepackage{subcaption}
\colorlet{punct}{red!60!black}
\definecolor{background}{HTML}{EEEEEE}
\definecolor{delim}{RGB}{20,105,176}
\colorlet{numb}{magenta!60!black}

\usepackage{color}
\definecolor{deepblue}{rgb}{0,0,0.5}
\definecolor{deepred}{rgb}{0.6,0,0}
\definecolor{deepgreen}{rgb}{0,0.5,0}
\definecolor{mygray}{rgb}{0.5,0.5,0.5}

\DeclareFixedFont{\ttb}{T1}{txtt}{bx}{n}{8} % for bold
\DeclareFixedFont{\ttm}{T1}{txtt}{m}{n}{12}  % for normal

\newcommand\pythonstyle{\lstset{
	language=Python,
	basicstyle=\linespread{0.8}\footnotesize,
	otherkeywords={self, None, True, False},             % Add keywords here
	keywordstyle=\ttb\color{deepblue},
	emph={MyClass,__init__},          % Custom highlighting
	emphstyle=\ttb\color{deepred},    % Custom highlighting style
	stringstyle=\color{deepgreen},
	frame=tb,                         % Any extra options here
	showstringspaces=false,            % 
	upquote=true,
    columns=fullflexible,
    breaklines=true,
    postbreak=\mbox{\textcolor{red}{$\hookrightarrow$}\space}, 
    numbers=left,                    % where to put the line-numbers; possible values are (none, left, right)
    xleftmargin=1.5em,
  numbersep=1pt,                   % how far the line-numbers are from the code
  numberstyle=\tiny\color{mygray}, % the style that is used for the line-numbers
  showtabs=false,                  % show tabs within strings adding particular underscores
  stepnumber=1,                  
  escapeinside={<@}{@>},
}}

\newcommand\jsonstyle{\lstset{
	string=[s]{"}{"},
    stringstyle=\color{blue},
    comment=[l]{:},
    commentstyle=\color{black},
    basicstyle=\linespread{0.5}\footnotesize,
    frame=tb,
    upquote=true,
    columns=fullflexible
}}

\lstnewenvironment{python}[1][]
{
\pythonstyle
\lstset{#1}
}
{}

\lstnewenvironment{json}[1][]
{
\jsonstyle
\lstset{#1}
}
{}

\newcommand\pythoninline[1]{{\pythonstyle\lstinline!#1!}}

\newcommand\jsoninline[1]{{\jsonstyle\lstinline!#1!}}

\newcolumntype{?}{!{\vrule width 1pt}}
  
\newcolumntype{L}[1]{>{\raggedright\let\newline\\\arraybackslash\hspace{0pt}}m{#1}}
\newcolumntype{C}[1]{>{\centering\let\newline\\\arraybackslash\hspace{0pt}}m{#1}}
\newcolumntype{R}[1]{>{\raggedleft\let\newline\\\arraybackslash\hspace{0pt}}m{#1}}
	
\graphicspath{{figures/}}

\newcommand{\zeynep}[1]{{\color{black}#1}}

\providecommand{\keywords}[1]{\textbf{\textbf{Keywords}} #1}
\usepackage[square,numbers]{natbib}
\begin{document}

%\title{AccaSim: a Scalable and Customazible {\normalfont H}PC {\normalfont Sim}ulator \\ for Workload Management} 

\title{AccaSim: a Customizable Workload Management Simulator for Job Dispatching Research in HPC Systems}

\author{
  Cristian Galleguillos\footnote{Pontificia Universidad Cat\'olica de Valpara\'iso, 2362807 Valpara\'iso, Chile.} $^{,}$\footnote{University of Bologna, 40126 Bologna, Italy} \and
  Zeynep Kiziltan$^{\dagger}$ \and 
  Alessio Netti$^{\dagger}$ \and
  Ricardo Soto$^{*}$
 }
% \institute{
% 	Cristian Galleguillos \and Ricardo Soto \at Pontificia Universidad Cat\'olica de Valpara\'iso, 2362807 Valpara\'iso, Chile
%     \\\email{cristian.galleguillos.m@mail.pucv.cl}%, ricardo.soto@pucv.cl}
%     \and
%     Cristian Galleguillos \and Zeynep Kiziltan \and Alessio Netti \at University of Bologna, 40126 Bologna, Italy
%  %   \\\email{\{zeynep.kiziltan@unibo.it, alessio.netti}@unibo.it}
% }

\maketitle

\input{sections/abstract.tex}

\input{sections/introduction.tex}

\input{sections/workload_manager.tex}

\input{sections/architecture.tex}

\input{sections/using.tex}

\input{sections/rel-work.tex}

\input{sections/sim_comparison.tex}

\input{sections/case_study.tex}

\input{sections/conclusions.tex}

\section*{Acknowledgements}
C. Galleguillos is supported by Postgraduate Grant PUCV 2018. A. Netti is supported by a research fellowship from the \textit{Oprecomp-Open Transprecision Computing} project. R. Soto is supported by Grant CONICYT/FONDECYT/ REGULAR/1160455. 
\zeynep{We are grateful to {\AA}ke Sandgren, Motoyoshi Kurokawa, and the Czech National Grid Infrastructure MetaCentrum, for providing, respectively, the Seth, RICC and the MetaCentrum workload datasets.}  We thank Alina S\^irbu for fruitful discussions on the work presented here. \zeynep{Finally, we appreciate the precious comments of the reviewers which helped improve the paper significantly. We especially thank Millian Poquet for signing his review and giving us the possibility to interact during the revision of the paper.}

\bibliography{main}

% The biographies and photos are required in the final version of the accepted paper
% \input{sections/bio.tex}

\end{document}

%% file: sections/abstract.tex
\begin{abstract}
%We present AccaSim, \sout{an HPC simulator for workload management} \zeynep{a simulator for workload management in HPC systems}. Thanks to \zeynep{AccaSim's scalability to large workload datasets and support for easy customization} \sout{and high customizability features of AccaSim}, as well as the practical automated tools that it provides to aid experimentation, users can easily represent various real HPC systems \sout{resources}, develop \zeynep{novel advanced dispatchers}, and \sout{carry out  experiments}   \zeynep{test and evaluate} them in a convenient way across different workload sources. AccaSim is thus an attractive tool for conducting \sout{controlled experiments in HPC dispatching research.} job dispatching research in HPC systems. 

\zeynep{We present AccaSim, a simulator for workload management in HPC systems. Thanks to AccaSim's scalability to large workload datasets, support for easy customization, and practical automated tools to aid experimentation, users can easily represent various real HPC systems, develop novel advanced dispatchers and evaluate them in a convenient way across different workload sources. AccaSim is thus an attractive tool for conducting job dispatching research in HPC systems. }

\end{abstract}

\keywords{HPC systems, workload management system, job dispatching problem, simulation tool, dispatcher development, dispatcher evaluation}

%% file: sections/introduction.tex
\section{Introduction}

High Performance Computing (HPC) systems have become fundamental tools to solve complex, compute-intensive, and data-intensive problems in diverse engineering, business and scientific fields, enabling new scientific discoveries, innovation of more reliable and efficient products and services, and new insights in an increasingly data-dependent world. This can be witnessed for instance in the annual reports\footnote{\url{http://www.prace-ri.eu/praceannualreports/}} of PRACE and the recent report\footnote{\url{http://www2.itif.org/2016-high-performance-computing.pdf}} by ITIF which accounts for the importance of HPC to the global economic competitiveness.
 
As the demand for HPC technology continues to grow, a typical HPC system receives a large number of variable requests \zeynep{(jobs)} by its end users. This calls for the efficient management of the submitted workload and system resources. This critical task  is carried out by the  \emph{Workload Management System} (WMS) software component. Central to WMS is the \emph{dispatcher} which has the key role of deciding when and on which resources to execute the individual jobs  by ensuring high system utilization and performance. An optimal dispatching decision is a hard problem \cite{DBLP:journals/dam/BlazewiczLK83}, and yet suboptimal decisions could have severe consequences, like wasted resources and/or exceptionally delayed requests.  Efficient \zeynep{job} dispatching \zeynep{in an HPC system} is thus an active research area, see for instance \cite{DBLP:journals/tpds/BridiBLMB16} fo r an overview.

One of the challenges of \zeynep{job} dispatching research is the \zeynep{intensive} experimentation necessary for evaluating and comparing various \zeynep{dispatchers} in a controlled environment. The experiments differ  under a range of conditions with respect to the workload, the number and the heterogeneity of resources, and the dispatching  \zeynep{algorithms}. Using a real HPC system for experiments is not realistic for the following reasons. First, researchers may not have access to a real system. Second, it is impossible to modify the hardware components of a system, and often unlikely to access its WMS for any type of alterations. And finally, even with a real system permitting modifications in its WMS, it is inconceivable to ensure that distinct \zeynep{dispatchers} process the same workload, which hinders fair comparison. Therefore, simulating a WMS %\sout{in a synthetic HPC system} 
is essential for conducting controlled dispatching experiments. %\sout{Unfortunately, currently available simulators are not flexible enough to render customization in many aspects, limiting the scope of their usage.}

The contribution of this paper is the design and implementation of AccaSim, a WMS simulator developed for job dispatching research in HPC systems.  AccaSim is an open source, freely available library for Python, thus compatible with any major operating system,  and executable on a wide range of computers thanks to its lightweight installation and light memory footprint.  AccaSim is scalable to large workload datasets and provides support for easy customization, allowing to carry out experiments across different workload sources, resource types, and dispatching algorithms. Moreover, AccaSim enables users to \zeynep{develop} novel advanced dispatchers by exploiting information regarding the current system status, which can be extended for including custom behaviors such as power and energy consumption and failures of resources. Furthermore, AccaSim  aids users in their experiments via automated tools to generate synthetic workload datasets, to run simulation experiments and to produce plots to evaluate dispatchers. The researchers can thus use AccaSim to mimic any real system, \zeynep{including those possessing heterogeneous resources}, 
%\sout{by setting up the synthetic resources suitably}, 
develop advanced dispatchers using for instance power and energy-aware, fault-resilient algorithms, and test and \zeynep{evaluate} them in a convenient way over a wide range of  workload sources by using real workload traces or by generating them. 
%\sout{As such, AccaSim is an attractive tool for \sout{developing dispatchers and conducting controlled experiments in job dispatching research in HPC systems. } conducting job dispatching research in HPC systems. }

This paper extends an earlier version \cite{DBLP:conf/carla/GalleguillosKN17} by providing three new automated tools for workload generation, experimentation and plot generation, as well as a detailed comparison to the relevant existing simulators. In the rest of the paper, after giving the background in Section~\ref{sec:wms} on WMS in HPC \zeynep{systems}, we introduce the architecture and the main features of AccaSim in Section~\ref{sec:accasim}. 
We briefly describe AccaSim's implementation and customization, and show its various instantiations  in Section~\ref{sec:arc-implementation}. We discuss in Section~\ref{sec:related-work} the related work and contrast AccaSim in Section~\ref{sec:comparison} against the existing relevant simulators. We then present a case study in Section~\ref{sec:case-study} where we \zeynep{showcase} AccaSim's use in job dispatching research. We conclude in Section~\ref{sec:conclusion}.

%% file: sections/workload_manager.tex
\section{Workload Management \zeynep{in HPC Systems} }
\label{sec:wms}

\begin{figure}[!t]
 \centering
 \includegraphics[width=0.7\columnwidth]{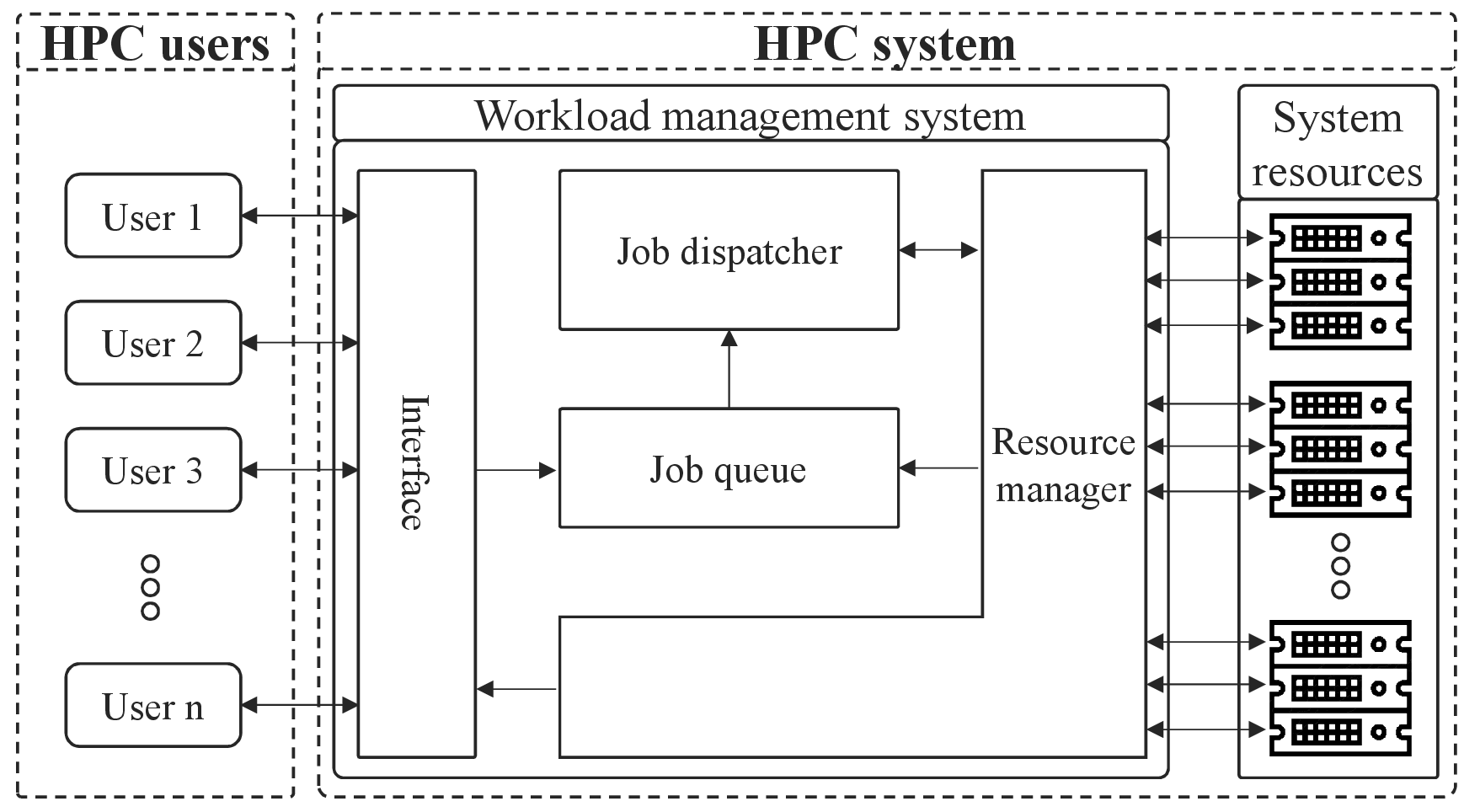}
 \caption{\zeynep{WMS in an HPC system.}}
 \label{fig:workload}
\end{figure}

A WMS is an important software of an HPC system, being the main access for the users to exploit the available resources for computing. A WMS manages user requests and the system resources through critical services.  A user request consists of the execution of a computational application over the system resources. Such a request is referred to as job and the set of all jobs are known as workload. The jobs are tracked by the WMS during all their states, i.e. from their submission time, to queuing, running, and completion. Once a job is completed, the results are communicated to the respective user. Figure~\ref{fig:workload} depicts a general scheme of a WMS. 

A WMS offers distinct ways to users for \emph{job submission} such as a GUI  and/or a command line interface. A submitted job includes the executable of a computational application, its respective arguments, input files, and the resource requirements. An HPC system periodically receives job submissions. Some jobs may have the same computational application with different arguments and input files, referring to the different running conditions of the application in development, debugging and production environments. When a job is submitted, it is placed in a \emph{queue} together with the other pending jobs (if there are any). The time interval during which  a job remains in the queue is known as waiting time.
The queued jobs compete with each other to be executed on limited resources. 

A \emph{job dispatcher} decides which jobs waiting in the queue to run next (\textit{scheduling}) and on which resources to run them (\textit{allocation}) by ensuring high system utilization and performance. The dispatching decision is generated according to a policy using the current system status, such as the queued jobs, the running jobs and the availability of the resources. A suboptimal dispatching decision could cause resource waste and/or exceptional delays in the queue, worsening the system performance and the perception of its users. A (near-)optimal dispatching decision is thus a critical aspect in a WMS. 

The dispatcher periodically communicates with a \emph{resource manager} of the WMS for obtaining the current system status. The \emph{resource manager} updates the system  status through a set of active monitors, one defined on each resource which primarily keeps track of the resource availability. The WMS systematically calls the \emph{dispatcher} for the jobs in the queue. An answer means that a set of jobs are ready for being executed. Then the dispatching decision is processed by the \emph{resource manager} by removing the ready jobs from the queue and sending them to their allocated resources. Once a job starts running, the \emph{resource manager} turns its state from ``queued'' to ``running''. The \emph{resource manager} commonly tracks the running jobs for giving to the WMS the ability to communicate their state to their users through the interface, and in a more advanced setting to (let the users) submit again their jobs in case of resource failures. 
When a job is completed, the \emph{resource manager} turns its state from ``running'' to ``completed'' and communicates its result to the interface to be retrieved by the user. 

%% file: sections/architecture.tex
\section{AccaSim Architecture and Main Features}\label{sec:accasim}

\begin{figure}[!t]
  \centering
  \includegraphics[width=0.8\columnwidth]{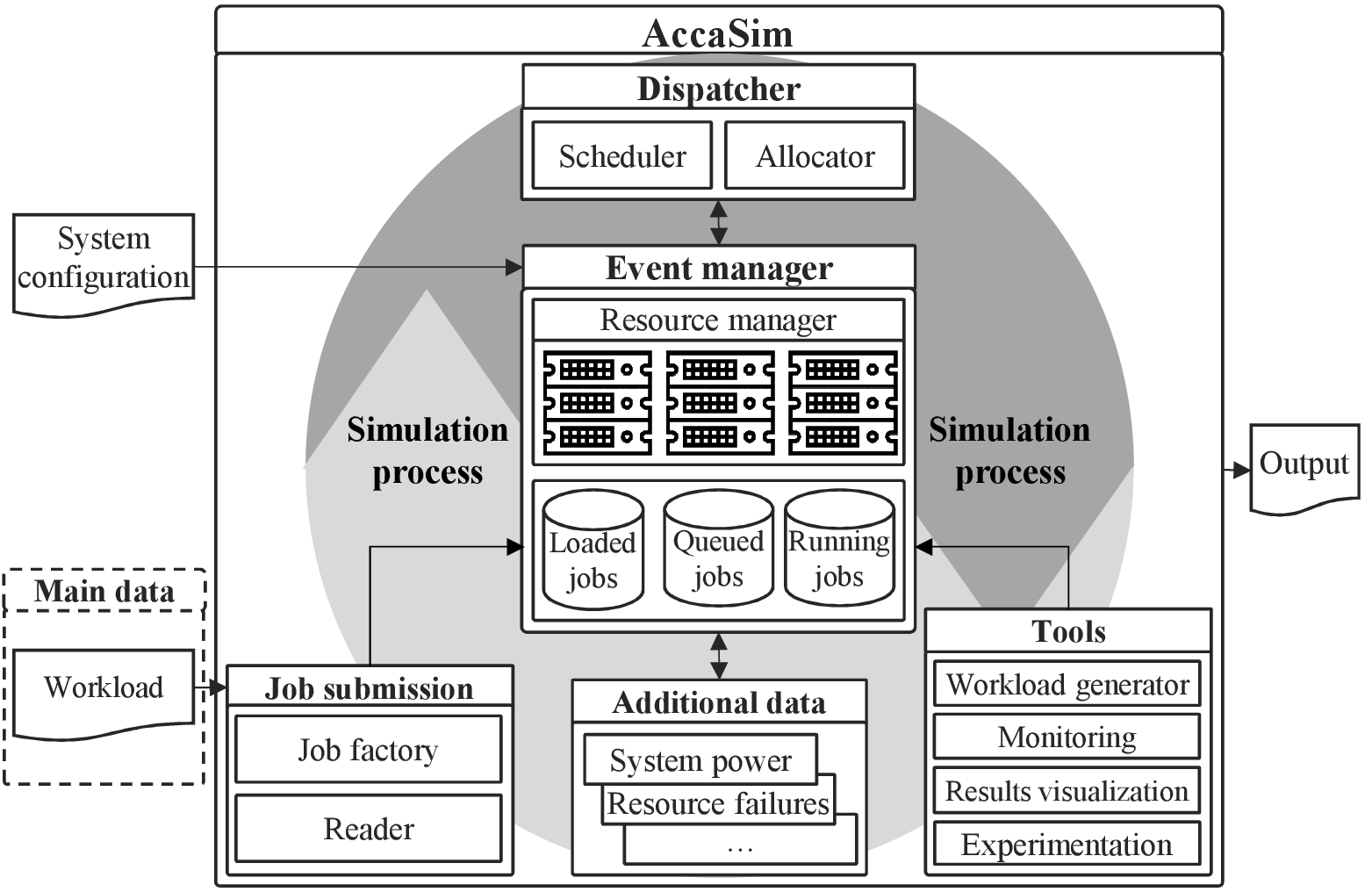}
  \caption{AccaSim architecture.}
  \label{fig:architecture}
\end{figure}

AccaSim enables to simulate the WMS of any real HPC system with minimum effort and facilitates  the study of various issues related to \zeynep{dispatchers}, such as feasibility, behavior, and performance, accelerating the dispatching research process. In this section, we present the architecture and \zeynep{highlight} the main features of AccaSim.  

AccaSim is designed as a discrete event simulator. The simulation is guided by certain events that belong to a real HPC system. These events are mainly collected from the workload and correspond to the job submission, starting and completion times, referred to as $T_{sb}$, $T_{st}$ and $T_{c}$, resp. The architecture of AccaSim is depicted in Figure~\ref{fig:architecture}. Since there are no real users for submitting jobs nor real resources for computation during simulation, the first step for starting a simulation is to define the synthetic system with its jobs and resources. 

\paragraph{Job submission.}
This component mimics the job submission of users. The main input data is the workload dataset provided in the form of a file which includes job descriptions.  The default \emph{reader} subcomponent reads the input file in Standard Workload Format (SWF)\cite{DBLP:journals/jpdc/FeitelsonTK14}  and passes the parsed data to the \emph{job factory} subcomponent for creating the synthetic jobs for simulation, keeping the information related to their identification, submission time, duration and request of system resources. The \emph{job factory} can extend this basic information with additional attributes for the synthetic jobs, such as \zeynep{job duration estimation}  which is a useful information for many dispatching algorithms  \cite{DBLP:conf/mod/GalleguillosSKB17}. The synthetic jobs are then mapped to the \emph{event manager} component, simulating the job submission process. The main data input is customizable in the sense that any workload dataset file can be used. This is possible thanks to the \emph{reader} which can be adapted easily to parse any workload  dataset file format. Consequently, AccaSim can be employed with any workload source  corresponding to an existing workload dataset or to a synthetic one produced by a workload generator. 

\paragraph{Event manager.}
This is the core component of the simulator, which mimics the behavior of the synthetic jobs and the presence of the synthetic resources, and manages the coordination between the two.  
Differently from a real WMS, the \emph{event manager} tracks the jobs during their artificial life-cycle by maintaining all their possible states ``loaded'', ``queued'', ``running'' and ``completed'' via \zeynep{certain} events. During simulation, at each time point $t$:

\begin{itemize}

\item the \emph{event manager} checks  if $t=T_{sb}$ for some jobs. 
If the submission time of a job is not yet reached, the \emph{event manager} assigns the  job the ``loaded'' state meaning in the real context that the job has not yet been submitted.  If instead the submission time of a job is reached, the \emph{event manager} updates its status to ``queued'';

\item  the \emph{dispatcher} component gives a dispatching decision on (the subset of) the queued jobs, assigning them an immediate starting time. The \emph{event manager} reveals that $t=T_{st}$ for some waiting jobs and consequently  updates their status to ``running''; 

\item 
the \emph{event manager} checks if $t=T_c$ for currently running jobs. Since these jobs were dispatched in a previous time point, their starting and completion times are known. \zeynep{The completion time of a job is  the sum of its starting time and duration, which are known from the workload data}. If the completion time of a job is reached, the \emph{event manager}  updates its status to ``completed''.
\end{itemize}
The \emph{resource manager} subcomponent of the \emph{event manager} defines the synthetic resources of the system using a system configuration file as input, and then mimics their allocation and release at the job starting and completion times. Hence, at a time point $t$, if a job starts, the \emph{resource manager} allocates for the job the resources decided by the \emph{dispatcher}; and if it completes, the \emph{resource manager} releases its resources. The system configuration file can be customized according to the needed types of resources, \zeynep{which renders the simulation of a system possessing heterogeneous resources} possible.  

AccaSim is designed to maintain a low consumption of memory for scalability \zeynep{to large workload datasets}, therefore job loading is performed in an incremental way, loading only the jobs that are near to be submitted at the corresponding simulation time, as opposed to loading them once and for all. Moreover, completed jobs are removed from the system so as to release space in the memory.

\paragraph{Dispatcher.} 
This component, responsible for generating a dispatching decision, interacts with the \emph{event manager} for retrieving the current system status regarding the queued jobs, the running jobs, and the availability of the resources. \zeynep{Note that the \emph{dispatcher} is not aware of job durations. This information is known only by the \emph{event manager} to  stop the jobs at their completion time in a simulated environment. Therefore, the dispatching decision can be solely based on job duration estimations which are supplied as a job attribute. This has no impact on the execution of jobs, which are always allowed to run for their entire duration, despite the presence of estimation errors.}
The \emph{scheduler} and the \emph{allocator} subcomponents of the \emph{dispatcher} are customizable according to the \zeynep{algorithms} of interest. Currently implemented and available \zeynep{schedulers} are: First In First Out (FIFO), Shortest Job First (SJF), Longest Job First (LJF) and Easy Backfilling with FIFO priority (EBF) \cite{DBLP:conf/cluster/WongG07}; and \zeynep{allocators are}:  First-Fit (FF) which allocates to the first available resource, and Best-Fit (BF) 
which sorts the resources by their current load (busy resources are preferred first), thus trying to fit as many jobs as possible on the same resource, to decrease the fragmentation of the system.  

\paragraph{Additional data.}
It has been shown in the last decade that system performance can be enhanced greatly if the  dispatchers  are aware of additional information regarding the current system status, such as energy and power consumption of the resources \cite{DBLP:conf/jsspp/ZhouLTD13,DBLP:conf/supercomputer/AuweterBBBHHPTW14,DBLP:conf/sc/BodasSRH14,DBLP:conf/cp/BorghesiCLMB15}, resource failures \cite{DBLP:conf/icpp/LiGLS07,DBLP:conf/ccgrid/BrandtDGMPTW08}, and the heating/cooling conditions \cite{DBLP:journals/tpds/TangGV08,BANERJEE2011134}. The \emph{additional data} component of AccaSim provides an interface to integrate such extra data to the system  which can then be utilized to develop and experiment with advanced dispatchers which are for instance energy and power-aware, fault-resilient and thermal-aware. The interface lets receive the necessary data externally from the user, make the necessary calculations together with some input from the \emph{event manager}, all customizable according to the need, and pass back the result to the \emph{event manager} so as to transfer it to the \emph{dispatcher}.  

\paragraph{Output.}
The output file contains two types of data. The first regards the execution of the dispatching decision for each job,  such as the starting time, the completion time and its resource allocation, which gets updated each time a job completes its execution.  This type of data can be utilized to contrast the quality of the dispatching decisions from different perspectives. An example is the effect on synthetic system resource utilization: how many and which resources are used in the system, and how they are distributed over the nodes.  Another example is the impact on system performance.  With the increasing trend in employing HPCs for real-time applications which cannot tolerate delays \cite{mohamed2014real}, some critical aspects of system performance are job response times and  system throughput. The second type of output data regards the simulation process, specifically the CPU time required by the simulation tasks like job loading, generation of the dispatching decision, and the total amount of memory used during simulation, which gets updated at each simulation time point. This type of data can be used, for instance, to evaluate the performance of the simulator, as well as the performance of the dispatchers in terms of the time they incur for generating a decision. 

\paragraph{Tools.}
The  \emph{tools} let users follow the simulation process and facilitate their dispatching experimentation. We will demonstrate their utility in Section \ref{sec:case-study}. The \emph{monitoring} subcomponent includes the \emph{system status} and \emph{system utilization} subcomponents. The \emph{system status} allows tracking the current system status, such as the number of queued jobs, the running jobs, the completed jobs, the availability of the resources, etc. The \emph{system utilization} instead shows in a GUI a representation of the allocation of resources by the running jobs during the simulation. 

The \emph{results visualization} subcomponent renders the automatic generation of different types of plots for evaluating the quality of dispatching decisions as well as  the performance of the dispatchers. 
The \emph{experimentation} subcomponent instead renders the automation of complex experiments. After configuring the simulator with a workload dataset, a system to simulate,  and a set of dispatchers, the \emph{experimentation} performs a simulation for each dispatcher and then produces comparative plots through the \emph{results visualization}.
 
When doing dispatching research with a real workload dataset, users could face issues such as the dependency on the real system configuration which hinders testing  with other system configurations, the small size of the  dataset preventing scalability tests, or the unavailability of certain data in the dataset for testing specific cases. To tackle this, AccaSim provides a \emph{workload generator} subcomponent which  produces a synthetic workload dataset. This subcomponent exploits the data contained in a real workload dataset by  mimicking, through statistical methods, its distributions for job submission times, jobs resource requests, and job durations. The generated dataset is  written to a  file in the SWF format. Other file formats can as well be considered by customizing its subcomponents.

To \zeynep{highlight} the \emph{main features}, \zeynep{(i) AccaSim is designed to be scalable to large workload datasets;} (ii) AccaSim is customizable in its workload source, resource types, and dispatching \zeynep{algorithms}, \zeynep{providing maximum flexibility in representing a WMS}; (iii) AccaSim enables users to develop novel advanced dispatchers by exploiting information regarding the current system status, which can be extended for including custom behaviors such as energy and power consumption and failures of the resources; (iv) Accasim provides output data and automated tools to analyze the results, to follow the simulation process and facilitate dispatching experimentation. 

%% file: sections/using.tex
\section{Implementation,  Customization, and Instantiation}\label{sec:arc-implementation}

\begin{figure}[!t]
 \centering
     \includegraphics[width=0.65\textwidth]{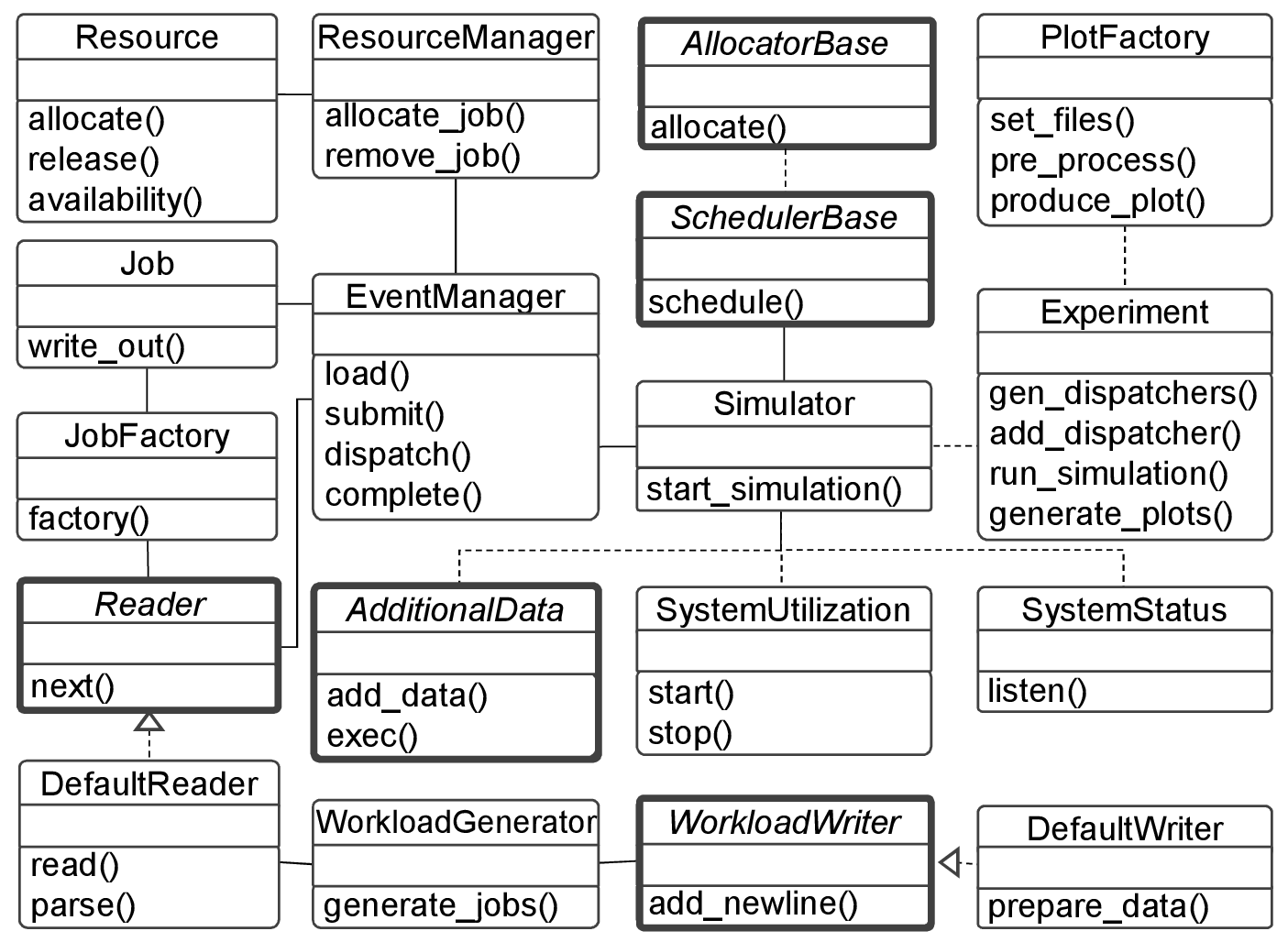}
 \caption{AccaSim class diagram.}\label{fig:class-diag}
\end{figure}

In this section, we briefly describe AccaSim's implementation and customization, and show its various instantiations. \zeynep{This not only serves to depict the internal organization of AccaSim, but also provides evidence on how easy it is to use and customize.}

AccaSim is implemented in Python which is an interpreted, object-oriented, high-level programming language, freely available for any major operating system, and is widely used in academia and industry.\footnote{\url{https://www.python.org/events/python-events/}} All the dependencies used by AccaSim are part of Python 3.5 and newer versions, except the matplotlib, scipy, sortedcontainters and psutil packages which can be easily installed using the pip management tool. The source code is available under MIT License, together with a documentation on the AccaSim website.\footnote{\url{http://accasim.readthedocs.io/en/latest/}} A release version is available as a package in the PyPi repository.\footnote{\url{https://pypi.org}} \zeynep{Customization} is driven by the abstract classes and the inheritance capabilities of Python. The UML class diagram of the main classes is shown in Figure~\ref{fig:class-diag} where the abstract classes associated to the customizable components are highlighted in bold.  

\paragraph{The simulator.}

A basic AccaSim instantiation is detailed in Figure \ref{fig:implementation}. A simulator object is created in line 11  by instantiating the \emph{Simulator} class. It receives as arguments a workload  dataset file in, for instance, SWF, a system configuration file in JSON format, and a dispatcher object, with which the synthetic system is generated and loaded with all the default features.

\begin{figure}[!t]
  \begin{python}
  from accasim.base.simulator_class import Simulator
  from accasim.base.scheduler_class import FirstInFirstOut
  from accasim.base.allocator_class import FirstFit
  from accasim.utils.plot_factory import PlotFactory

  workload = 'workload.swf'
  sys_cfg = 'sys_config.json'
  
  allocator = FirstFit()
  dispatcher = FirstInFirstOut(allocator)
  simulator = Simulator(workload, sys_cfg, dispatcher) 
  output_file = simulator.start_simulation()
  
  plot_factory = PlotFactory('decision', sys_cfg)
  plot_factory.set_files(output_file, 'my_plot')
  plot_factory.produce_plot('slowdown')
  \end{python}
\caption{A basic AccaSim instantiation.}
\label{fig:implementation}
\end{figure}

The workload  dataset file is handled by an implementation of the abstract  \emph{Reader} class, which is the SWF-based \emph{DefaultReader} by default. The file is read and parsed by the \emph{read()} and \emph{parse()} methods. By implementing the \emph{Reader} class appropriately, AccaSim can be customized to read any workload dataset  file format beyond SWF, or to read workloads from any source, not necessarily from a file. The system configuration file, which is processed by the \emph{ResourceManager} class, defines the synthetic resources. The file has two main contents. The first specifies the resource types and their quantity \zeynep{in a node belonging to a group}, which is useful for modeling \zeynep{HPC systems possessing heterogeneous resources}. The second, instead, defines the number of nodes of each group. See Figure \ref{fig:json-seth} for an example. The user is free to mimic any real system by customizing this configuration file suitably.

The dispatcher object is composed by implementations of the abstract \emph{SchedulerBase} and \emph{AllocatorBase} classes. Both classes must implement their main methods, \emph{schedule()} and \emph{allocate()} respectively, to deal with the scheduling and the allocation decisions of the dispatching. This illustrative instantiation exemplifies a specific instance of the \emph{Simulator} class, using as scheduler the \emph{FirstInFirstOut} class, which implements \emph{SchedulerBase} with FIFO, and as allocator the \emph{FirstFit} class, which implements \emph{AllocatorBase} using FF. Both the \emph{FirstInFirstOut} and \emph{FirstFit} classes are  available in the library  for importing, as done in lines 2-3 of Figure \ref{fig:implementation}. AccaSim can be customized in its dispatching \sout{method} \zeynep{algorithm} by implementing the abstract \emph{SchedulerBase} and \emph{AllocatorBase} classes as desired. 

In line 12, the \emph{start\_simulation()} method launches the simulation with the following optional arguments:

\begin{minipage}{\linewidth}
\begin{python}[frame=None,numbers=none]
    simulator.start_simulation(
    			system_status=True, 
			system_utilization=True, 
			additional_data=None)
\end{python}
\end{minipage}

\noindent
which serve to require the use of the \emph{system status}, the \emph{system utilization}, and the  \emph{additional data} tools of the simulator. The additional\_data argument is an array of objects where each object is an implementation of the abstract \emph{AdditionalData} class, giving the possibility to customization in terms of the extra data that the user may want to provide to the system for dispatching purposes. After the simulation is finished, the output data file is returned.

The last three lines in Figure \ref{fig:implementation} serve to use the automated plot generation tool. In line 14, the \emph{PlotFactory} class is instantiated using two arguments. The first indicates the plot type to be produced, as a decision-related or performance-related type. A decision-related plot shows metrics related to the quality of the dispatching decision, such as the job slowdown \cite{feitelson2001metrics} or queue size, while a performance-related plot serves to show metrics related to the performance of the dispatcher, such as the average CPU time at a simulation time point. Examples of such plots will be shown in Section \ref{sec:case-study}. The second argument is instead the system configuration file which is necessary for the resource specific plots. In line 15, the output file of the simulator is set to be analyzed through the \emph{set\_files()} method, together with a label to be used in the plots. Finally, the \emph{produce\_plot()} method produces the desired plot as specified in its argument.

\begin{figure}[!t]
  \begin{python}[firstnumber=5]
  [...]
  from accasim.base.scheduler_class import ShortestJobFirst
  from accasim.experimentation.experiment import Experiment
  
  experiment = Experiment('my_experiment', workload, sys_cfg)
  sched_list = [FirstInFirstOut, ShortestJobFirst]
  alloc_list = [FirstFit]
  experiment.gen_dispatchers(sched_list, alloc_list)
  experiment.run_simulation()
  \end{python}
\caption{An AccaSim instantiation using the experimentation tool.}
\label{fig:implementation-exp}
\end{figure}

\paragraph{The experimentation tool.}
In Figure \ref{fig:implementation-exp}, an AccaSim instantiation that uses the  \emph{experimentation} tool is detailed. The first 4 lines related to  imports and assignment statements are the same as lines 2, 3, 6 and 7 in Figure \ref{fig:implementation} and are therefore omitted. An experiment object is created in line 9 by instantiating the \emph{Experiment} class which takes as arguments the name of the experiment (which is used to name the output directory as well),  the workload  dataset file, and the system configuration file,  along with the the optional arguments supported by the \emph{Simulator} class. In line 12,  the dispatchers of interest are generated through the \emph{gen\_dispatchers()} method, which accepts as arguments a list of scheduler and allocator classes. In this  illustrative instantiation of the \emph{Experiment} class, we use the \emph{FirstInFirstOut} and the \emph{ShortestJobFirst} classes which implement FIFO and SJF scheduling, as well as the \emph{FirstFit} class which implements the FF allocation. All these classes are available in the library  for importing, as done in lines 6-7 of Figure \ref{fig:implementation-exp}. The \emph{gen\_dispatchers()} method then automatically creates the dispatchers corresponding to all possible combinations between the schedulers and the allocators, facilitating greatly the conduction of experiments on large sets of dispatchers. If users wish to experiment with a specific dispatcher, it can be formed by instantiating the corresponding implementation of \emph{SchedulerBase} and then passing the object to the \emph{add\_dispatcher()} method, similarly to what we have shown in the lines 9-11 in Figure \ref{fig:implementation} when instantiating the \emph{Simulator} class. Finally in line 13, the experiment is launched with the \emph{run\_simulation()} method which performs simulations for all configured dispatchers and produces all the available plots.

\paragraph{The workload generator tool.} The workload  dataset file can refer to a real workload dataset extracted from an HPC system, or to a synthetic one generated through an external workload generator such as AccaSim's own \emph{workload generator} tool. Figure \ref{fig:wgen_instantiation} shows its basic instantiation. A generator object is created in line 8 via the \emph{WorkloadGenerator} class which is available in the library  for importing, as done in line 1. It receives as arguments a real workload  dataset file to be mimicked, a system configuration file, and variables regarding performance and request limits. The performance variable is a dictionary storing the performance of each processing unit as a unit-value pair. The request\_limits variable instead defines the minimum and maximum request of each resource type available in the system. Finally, the jobs are generated in line 9  using the \emph{generate\_jobs()} method, which receives as arguments the number of jobs and the name of the output file in which the generated workload dataset is saved. 

\begin{figure}[!t]
  \begin{python}
  from accasim.experimentation.workload_generator import WorkloadGenerator
  
  workload = 'real_workload.swf'
  sys_cfg = 'sys_config.json'
  performance = {'core': 1.667}
  request_limits = {'min': {'core': 1, 'mem': 256}, 'max': {'core': 8, 'mem': 1024}}
  
  gen = WorkloadGenerator(workload, sys_cfg, performance, request_limits)
  jobs = gen.generate_jobs(500000, 'new_workload.swf')
  \end{python}
\caption{A basic workload generator instantiation.}
\label{fig:wgen_instantiation}
\end{figure}

As in the case of the simulator, the input workload dataset file is parsed by an implementation of the abstract \emph{Reader} class, which is \emph{DefaultReader} and implements an SWF reader by default. The output file is instead  written through an implementation of the abstract \emph{WorkloadWriter} class, which is the SWF-based \emph{DefaultWriter} by default. Similar to the \emph{Reader}, the output file format can be customized by implementing the \emph{WorkloadWriter} suitably. It is also possible to customize the job generation process via the optional arguments of the \emph{WorkloadGenerator} constructor, as detailed in the AccaSim documentation.

%% file: sections/rel-work.tex
\section{Related Work}\label{sec:related-work}

HPC systems have been simulated from distinct perspectives, for instance to model their network topologies~\cite{DBLP:conf/europar/AcunJBMCK15,DBLP:conf/sc/JainBWGK16,DBLP:journals/tpds/MubarakCRC17} or storage systems~\cite{DBLP:conf/sc/SnyderCLMRCBLBP15,DBLP:journals/tjs/NunezFGGC10}. There also exist simulators dealing with the duties of a WMS, as in our work, which are mainly focused on job submission, resource management and job dispatching. 

To the best of our knowledge, \zeynep{
%\sout{the most recent WMS simulator is presented in \cite{gonzalo2017scsf}} 
the WMS simulators most similar to AccaSim are %the Scheduling Simulation Framework (ScSF), 
ScSF,  %the Batch Scheduler Simulator (Batsim) 
Batsim, and %the GridSim based Job Scheduling Simulator (ALEA). 
Alea. The ScSF simulator \cite{gonzalo2017scsf}} emulates a real WMS, Slurm Workload Manager\footnote{\url{https://slurm.schedmd.com/}}, which is popular in many HPC systems. In \cite{lucero2011simulation,trofinoff2015mb} Slurm is modified to provide synthetic job submission, resource management and job dispatching through distinct daemons which run in diverse virtual machines and which communicate over RPC calls, and a dedicated simulator is implemented. ScSF extends this simulator with  automatic generation of synthetic job descriptions based on statistical data, but does not give the possibility to read real workload  datasets. The dependency on a specific WMS \zeynep{complicates} the customization, and together with the additional dependency on virtual Machines and MySQL, the  set up  of ScSF is rather complex. Moreover, ScSF requires  a significant amount  of resources in the machines where the simulation will be executed.  

% \begin{figure}[!t]
% 	\centering
% 	\includegraphics[width=.48\textwidth,trim=17 7 50 43,clip=true]{figures/Fig18.eps}
% 	\caption{RICC workload dataset.}
% 	\label{workload:flop-ricc}
% \end{figure}

Batsim \cite{DBLP:conf/jsspp/DutotMPR16} is developed on top of the SimGrid simulation framework.\footnote{\url{http://simgrid.gforge.inria.fr/}} Batsim decouples the dispatcher from the simulator  and allows it to be implemented in any programming language, yet both the simulator's and the dispatcher's source code and binaries are available only for GNU/Linux.  Batsim takes as input a file in a JSON-based format, %Batsim also provides a script to translate an input file in the SWF format to the JSON-based format,  allowing to read thus real workload datasets.  
and provides a script to translate from SWF  %to the JSON-based format, 
with which it is possible to read real workload datasets. %\alessio{Such script, as well as available dispatchers, allows modeling and dealing only with one resource type, such as processor cores. Moreover, while including additional attributes can be easily achieved, modeling multiple resource types requires a significant effort, when users wish to model a system using heterogeneous resources. In fact, although SimGrid supports a heterogeneous resource type definition, the concept of a single node possessing heterogeneous resources is not natively implemented in the simulator. This calls for a more advanced job modeling to be implemented by users, and such changes must be reflected both in the system model and dispatchers. }
%However, some information regarding jobs' resource requests is lost during translation, because the JSON-based format restricts jobs' requests to the number of nodes necessary to cover their core or processor requests, as opposed to the necessary number of cores or processors, amount of memory etc.  Consequently, as soon as a Batsim dispatcher allocates a node for a job, the node will not be available for other jobs, even if the node still has available cores or processors.  
%An AccaSim dispatcher, instead, considers all specific resource requests of a job detailed in the SWF format and allocates the nodes to their maximum capacity so as to reduce resource wastage.  
%Due to this difference, an AccaSim dispatcher algorithm may spend more time in allocation than the corresponding Batsim algorithm, at the same time may allocate more jobs waiting in the simulated system's queue and consequently reduce the makespan in the dispatching decision.
%In Batsim, 
However,  all jobs are loaded in memory at the beginning of simulation which can hinder the performance when experimenting with a large number of jobs. \zeynep{While users can define different resource types as supported by SimGrid, the concept of a single node possessing heterogeneous resources is not natively implemented in the simulator. This calls for significant effort when users wish to model a system using heterogeneous resources. The dispatchers need to be adapted as well in order to take into account the new representation of a system.}
%\sout{Moreover, job attributes are fixed and are not customizable so as to exploit relevant information about jobs (such as job duration estimation) for dispatching purposes. Although SimGrid allows to define different resource types, Batsim and the dispatchers can deal only with jobs requesting cores or processors. Consequently, unlike AccaSim, Batsim cannot be used to simulate an heterogeneous HPC system.}
Similar to AccaSim, additional data regarding the current system status can be used in Batsim for instance, to model the energy consumption of the system. The type of data, however, depends exclusively on the capabilities of SimGrid. And finally, while Batsim includes a  workload generator, it is simple, useful for testing purposes only, and is not intended for dispatching research.

Alea~\cite{DBLP:conf/simutools/KlusacekR10} is developed on top of the GridSim simulation framework.\footnote{\url{http://www.cloudbus.org/gridsim/}} \zeynep{ %However, users can implement new dispatchers through the available interfaces. 
Job submission, resource management and job dispatching are driven by the predefined workload format, resource types, and dispatchers. The implementation in Java  is open-source and cross-platform. However, any customization to the simulator needs to be done at the source code level, which can be complicated and error-prone.} 
%There also exist %Python Scheduler Simulator (
PYSS~\cite{DBLP:conf/sc/LiuW15,DBLP:conf/jsspp/LelongRT17,DBLP:journals/tpds/MuraliV18} and %Ocaml EASY-Backfilling Simulator (
OCS~\cite{DBLP:conf/sc/GaussierGRT15} have similar characteristics to Alea,  but provide less advanced WMS features as they are developed primarily for a specific research work in dispatching. In general, simple simulators like PYSS and OCS hinder the design of novel advanced dispatchers and their evaluation which requires a more flexible way to represent a WMS.

In \cite{DBLP:journals/concurrency/Gomez-MartinV015}, an energy aware WMS simulator, called Performance and Energy Aware Scheduling (PEAS) simulator is described. With the main aim being to minimize the energy consumption and to increase the throughput of the system, PEAS uses predefined dispatchers and workload dataset file format, and the system power calculations are based on fixed data from SPEC benchmark\footnote{\url{https://www.spec.org/power_ssj2008/}} considering the entire processor at its max load. PEAS is available only as  GNU/Linux binary, therefore  it is not customizable in any of these aspects.   

Brennan et al.~\cite{DBLP:conf/dsrt/BrennanKH14} define a framework for WMS simulation, called Cluster Discrete Event Simulator (CDES), which uses predefined scheduling algorithms and relies on specific resource types. Although CDES allows reading real workload datasets for job submission, it loads all jobs in memory at the beginning of the simulation, like Batsim does. Moreover, the implementation is not available which prevents any form of customization.    

In \cite{DBLP:conf/alar/HurstRLH10}, a WMS simulator based on a discrete event library called Omnet$^{++}$\footnote{\url{http://www.omnetpp.org/}} is introduced. Similar to ScSF,  only automatically generated synthetic job descriptions are accepted for job submission. Since Omnet++ is primarily used for building network simulators and is not devoted to workload management, there exist issues such as the inability to consider different types of resources as in  CDES. Moreover, due to lack of documentation, it is hard to understand to what extent the simulator is customizable. 

%\cristian{Similar to already presented WMS simulators but with less advanced features some specific HPC dispatcher simulators have been developed to carry out research in dispatchers ~\cite{DBLP:conf/sc/GaussierGRT15,DBLP:conf/sc/LiuW15,DBLP:conf/jsspp/LelongRT17,DBLP:journals/tpds/MuraliV18}, such as Python Scheduler Simulator (PYSS) or Ocaml EASY-Backfilling simulator (OCS). Those simulators are characterized to be simple and lightweight, in exchange, they offer a limited system modeling and job features. Although their source code is available, introducing minor changes may require major effort. %, where only processors are defined as system resources, and jobs are represented in terms their requested wall-times and number of cores. Although their source code is available to customize job attributes or dispatching algorithms, doing minor changes, such as including an extra resource type, may require an extra effort since their implementations were not intended to be a software library/framework, which could ease the customization of the simulation. 

The main issues presented in the existing  WMS simulators w.r.t. to AccaSim can be summarized as complex set up and need of many virtual machines and resources, inflexibility in the workload source and resource types, limited support for additional data, potential performance degrade with large workload datasets, \zeynep{difficulty or the impossibility} of the customization of the WMS, platform restriction, and unavailable or undocumented implementation. \zeynep{As AccaSim is developed for facilitating job dispatching research in HPC systems, it is designed to be scalable to large workload datasets and provides maximum flexibility in representing a WMS in terms of workload source, resource types, and dispatchers. It is open-source and cross-platform, simple to install and use, and is easy to customize via abstract class implementations without having to touch the source code. }

%% file: sections/sim_comparison.tex
\section{Comparison of Simulators}\label{sec:comparison}

In this section, we contrast AccaSim  with a critical attention against ScSF, Batsim and Alea which are the most similar simulators to AccaSim. 

\subsection{Comparison to ScSF}

ScSF\footnote{\url{http://frieda.lbl.gov/download}} is a complex framework which needs an entire testing environment for running. The environment should have at least two real or virtual machines with dedicated resources, enough hard disk space for the simulator and its components, and external applications such as a  database. The network connection is also a key point in the simulation, since it is required to have a low latency in order to maintain a fast link between its components.  We do not compare AccaSim to ScSF experimentally for the following reasons. First, the physical resources needed for experimentation with ScSF are much more than those required by AccaSim. Second, the processes involved in a simulation are more complicated, and they are not encapsulated in a single parent process, as in AccaSim, which hinders a  fair comparison. For instance, there are processes that are executed in the MySQL database or that depend on ssh connections, which can affect the performance evaluation. Third, job submission in ScSF is performed only by its own workload generator which restricts the experiments to the synthetic jobs generated by ScSF itself. 
%\sout{The generated job attributes are identification, name, the number of required cores and duration, and exclude some essential attributes required by dispatching algorithm such as the memory request, which is useful for conducting  allocation in a manner similar to a real system, or the estimated job duration, which is used in the majority of the schedulers to produce robust scheduling plans.}

\subsection{Comparison to Batsim and Alea} 
\zeynep{

We here conduct an experimental study to compare the performance of AccaSim to Batsim and Alea using three real workload datasets, which are freely available in SWF. 
%In the following, we first \sout{give some directions about the required steps to use the simulators under comparison}\zeynep{underline some issues around the use of }. Then, we describe the workload datasets and the experimental setup. Finally, we report our experimental results. 
The study is performed on an Ubuntu 16.04 machine with an Intel Core i7-2600 CPU, 16 GB of RAM and a WD10EZEX HDD with 1 TB of capacity. The software used for each simulator experiment are AccaSim 1.0 with Python 3.6.5, Batsim 2.0.0 with Batsched 1.2.0, and finally Alea 4.0 with OpenJDK 1.8.0\_171 and 4 GB of max. heap size. All the scripts used to setup and run to experiments, and to evaluate their results are available on the AccaSim GitHub repository.\footnote{\url{https://github.com/cgalleguillosm/accasim/tree/journal/extra/journal_scripts/simulator_comparison}}

\paragraph{Workload datasets} 

It is important to compare the simulators' performance on datasets diverse in terms of size and time span, so as to derive robust conclusions on their behavior, especially on how they scale up to large workload datasets. The three datasets on which the experiments are based differ in these aspects. They range from medium-size to very large-size, and they are created in time periods ranging from a decade ago to recent years. The first dataset is based on a workload trace collected from the Seth cluster\footnote{\url{https://www.hpc2n.umu.se/resources/hardware/seth}} which was part of the High Performance Computing Center North of the Swedish National Infrastructure for Computing. The dataset file is available on-line\footnote{\url{http://www.cs.huji.ac.il/labs/parallel/workload/l_hpc2n/index.html}} in SWF, and it contains 202,871 jobs spanning through 4 years, from July 2002 to January 2006. Seth was composed of 120 nodes, 
%\cristian{\sout{each node with two AMD Athlon MP2000+ dual core processors with 1.667 GHz and 1 GB of RAM.} 
480 cores and 120 GB of RAM in total.

The workload trace on which the second dataset is based is collected from the RICC supercomputer \cite{DBLP:conf/ic-nc/Nakata11} which was part of RIKEN, an independent scientific research and technology institution of the Japanese government. The dataset file is available on-line\footnote{\url{http://www.cs.huji.ac.il/labs/parallel/workload/l_ricc/index.html}} SWF, and it contains 447,794 jobs spanning through 5 months, from May 2010 to September 2010. RICC was a massively parallel cluster, which was composed of 1,024 nodes, %\cristian{\sout{each node with two Intel Xeon 5570 CPUs quad core processor with 2.93GHz and 12 GB of RAM.} 
8192 cores and 12 TB of RAM in total.

The last workload dataset is based on a workload trace collected from the MetaCentrum Czech National Grid~\cite{DBLP:conf/jsspp/KlusacekTP15}. The dataset file is available on-line\footnote{\url{http://www.cs.huji.ac.il/labs/parallel/workload/l_metacentrum2/index.html}} in SWF, and it contains 5,731,100 jobs spanning through 2 years, from Jan 2013 to Apr 2015. The MetaCentrum grid \footnote{\url{https://metavo.metacentrum.cz/en/index.html}} is composed of several clusters, the composition of which has changed over the time. During the recorded period, it was composed of 19 clusters with 495 nodes, 8412 cores and 10 TB of RAM in total.

%with Seth being medium-size, RICC large-size, and MetaCentrum very large-size 
%This allows for a diverse experimental setup, and to characterize the simulators' performance and scalability.

%The two datasets differ in many aspects. The Seth has almost half of the number of jobs of the RICC. Regarding the  number of jobs submitted to the system during the recorded periods, the Seth has an average monthly submission of 4,779 jobs while the RICC has  89,558. The average CPU time requested for the Seth and RICC are 42.3 and 59.7 hours, with an average interarrival job submission frequency of 636 and 67 secs, respectively. Based on these, we understand that RICC was a system which worked at higher levels of utilization  and probably maintained a big queue size for longer periods. 

\paragraph{Experimental setup} 

Each experiment corresponds to the simulation of one of the three workload datasets using one of the three simulators. %and is repeated 10 times so as to obtain reliable and representative results. 
In order to isolate the core actions of a simulator from external factors, such as non-optimal dispatcher implementations, we use a dispatcher which rejects any submitted job.  While the rejecting dispatcher is available in AccaSim and Batsim, we implemented it ourselves in Alea. We evaluate the simulators' performance in terms of the  total CPU time required to run an experiment and memory footprint. To do so, we use a script which sequentially runs each experiment and repeats it 10 times as a child program in a new process so as to obtain reliable and representative results. The script records each experiment's start and ending time, and gathers the memory consumption every 10ms by using the Python psutil library.\footnote{\url{https://pypi.org/project/psutil/}} %and measure the total CPU time and memory consumption for all experiments under the same conditions. The script 

Batsim\footnote{\url{https://github.com/oar-team/batsim}} is conveniently packaged in the Nix package manager for an easy and clean installation on any Linux distribution with superuser privileges. Batsim does not accept SWF in input, and instead provides a script to convert SWF into the required format. This script also works as a workload preprocessor which removes jobs with incomplete or erroneous data. The CPU time and memory consumption of this preprocessing phase is not considered in the Batsim performance result. Instead in AccaSim and Alea, a similar preprocessing is carried out during job submission, therefore the corresponding CPU time and memory consumption are included in the AccaSim and Alea performance results.   
%, although in AccaSim this task is carried out during the simulation process by the dispatcher \alessio{and job factory components} like in a real context.

Alea\footnote{\url{https://github.com/aleasimulator/alea/}} is distributed as a Netbeans Java project in which the entire source code is available. All dependencies and a sample simulation configuration are provided. As opposed to Batsim, Alea accepts SWF in input. %The main drawback during the job reading is the need of specifying the number of jobs expected to read during a simulation. This number is defined in the experiment configuration, and it has to be satisfied during a simulation otherwise the reading procedure will crash. We used as a workaround the definition of a lower number of expected jobs to be read in comparison to available jobs in the workload, specifically for the Seth workload, setting 200,500 instead of 202,871 expected jobs to read. 
However, Alea needs the number of expected jobs in the simulation. Since the number of jobs in the workload may reduce during the preprocessing step, a mismatch with the workload size may crash the job submission process. We indeed faced the problem with the Seth dataset and worked around it by using a number of jobs (200,500), obtained by trial and error, lower than the size of the workload (202,871). Another issue in  Alea is that it includes hardcoded instructions for specific datasets or systems which may have to be modified for recent or custom datasets. This kind of implementation makes Alea rather difficult to use. %and requires users to modify the code of the simulator itself.

\paragraph{Experimental results}

\begin{table}[!t]
\centering
\begin{tabular}{|c|c|c|c|r|r|r|}
\hline
\multicolumn{4}{|c|}{\multirow{2}{*}{Workload}}                              & \multicolumn{3}{c|}{Simulator}                                                            \\ \cline{5-7} 
\multicolumn{4}{|c|}{}                                                        & \multicolumn{1}{c|}{AccaSim} & \multicolumn{1}{c|}{Batsim} & \multicolumn{1}{c|}{Alea} \\ \hline
\multirow{6}{*}{Seth} & \multicolumn{2}{c|}{Total time} & \(\mu\) & \textbf{00:15} & 00:34 & \textbf{00:15} \\ \cline{4-7} 
& \multicolumn{2}{c|}{(MM:SS)} & \(\sigma\) & 0.2 & 0.5 & 0.5 \\ \cline{2-7} 
& \multirow{2}{*}{Mem.} & \multirow{2}{*}{Avg.} & \(\mu\) & \textbf{18} & 596 & 161 \\ \cline{4-7} 
& \multirow{3}{*}{(MB)} &  & \(\sigma\) & 0.1 & 2.5 & 5.4 \\ \cline{3-7} 
&  & \multirow{2}{*}{Max.} & \(\mu\) & \textbf{18} & 964 & 209 \\ \cline{4-7} 
&  &  & \(\sigma\) & 0.1 & 0.2 & 23.7 \\ \hline
\multirow{6}{*}{RICC} & \multicolumn{2}{c|}{Total time} & \(\mu\) & 00:27 & 01:03 & \textbf{00:24} \\ \cline{4-7} 
& \multicolumn{2}{c|}{(MM:SS)} & \(\sigma\) & 0.5 & 0.7 & 0.2 \\ \cline{2-7} 
& \multirow{2}{*}{Mem.} & \multirow{2}{*}{Avg.} & \(\mu\) & \textbf{21} & 1,220 & 162 \\ \cline{4-7} 
& \multirow{3}{*}{(MB)} &  & \(\sigma\) & 0.1 & 5.4 & 5.6 \\ \cline{3-7} 
&  & \multirow{2}{*}{Max.} & \(\mu\) & \textbf{26} & 2,072 & 272 \\ \cline{4-7} 
&  &  & \(\sigma\) & 0.1 & 0.1 & 52.3 \\ \hline
\multirow{6}{*}{MC} & \multicolumn{2}{c|}{Total time} & \(\mu\) & \textbf{06:23} & 29:29 & 09:08 \\ \cline{4-7} 
& \multicolumn{2}{c|}{(MM:SS)} & \(\sigma\) & 4.1 & 14.2 & 3.7 \\ \cline{2-7} 
& \multirow{2}{*}{Mem.} & \multirow{2}{*}{Avg.} & \(\mu\) & \textbf{19} & 12,647 & 195\\ \cline{4-7} 
& \multirow{3}{*}{(MB)} &  & \(\sigma\) & 0.1 & 137.2 & 17.4 \\ \cline{3-7} 
&  & \multirow{2}{*}{Max.} & \(\mu\) & \textbf{19} & 15,431 & 1,165 \\ \cline{4-7} 
&  &  & \(\sigma\) & 0.2 & 6.7 &  234.4 \\ \hline
\end{tabular}
\caption{Performance comparison of AccaSim, Batsim and Alea.}
\label{table:sim-comparison}
\end{table}

We present the results in Table \ref{table:sim-comparison}, where the MetaCentrum dataset is abbreviated as MC, the total CPU time spent in an experiment is expressed in MM:SS, and the memory usage is expressed with its average and maximum values in MB. The reported values of an experiment are aggregated across all the 10 iterations, and both mean (\(\mu\)) and standard deviation (\(\sigma\)) are shown. Across the same dataset and metric, the best results are indicated in bold. 

It is clear to see that AccaSim uses up much less memory than the other simulators due to its incremental job loading and job removal capability. This approach is shared by Alea which shows better performance than Batsim. As was discussed in Section \ref{sec:related-work}, Batsim loads in memory the preprocessed data from the workload at the beginning of the simulation, which clearly hinders the performance when experimenting with a large workload dataset. As for the total CPU time, AccaSim and Alea show competitive results. Despite AccaSim's more general and costly approach in creating synthetic jobs that can have additional attributes with respect to Alea, the results are close with the medium-size Seth and large-size RICC datasets. AccaSim shows the best results with the very large-size MetaCentrum dataset. Batsim's performance worsens, as the workload size increases. This can be explained by its high memory consumption. In general, when an application requires high amount of memory, the OS has to employ auxiliary data structures  at the expense of reduced  performance. 

We can conclude that, Accasim is scalable to large workload datasets, and overall it performs much better than the similar simulators Batsim and Alea.}

%% file: sections/case_study.tex
\section{Case Study}\label{sec:case-study}

\zeynep{In this section, we present a case study to illustrate\sout{'s} AccaSim use in job dispatching research. We here focus primarily on dispatcher evaluation %, and finally 
and synthetic workload generation. AccaSim can as well be used to develop advanced dispatchers, see \cite{DBLP:conf/mod/GalleguillosSKB17} for an example. We leave further examples of dispatcher development in AccaSim to future work.}

The experimental study conducted in this section is performed on a CentOS 7.3 machine with two Intel Xeon E5-2630 v3 CPUs, 128GB of RAM, using Python 3.6.5 and Accasim 1.0. All the scripts used to setup and run the experiments, and to evaluate their results are available on the AccaSim GitHub repository.\footnote{\url{https://github.com/cgalleguillosm/accasim/tree/journal/extra/journal_scripts/case_study}}

% \begin{figure}[!t]
%   \begin{json}
%   {
% 	"system_name": "RICC - RIKKEN",
% 	"start_time": 1272639895,
% 	"equivalence": {
% 		"processor": {
% 			"core": 1
% 		}
% 	},
% 	"groups": {
% 		"g0": {
% 			"core": 8, 
% 			"mem": 12000000
% 		}
% 	},
% 	"resources": {
% 		"g0": 1024
% 	}
%   }
%   \end{json}
%   \caption{System configuration of RICC.}
%   \label{fig:json-ricc}
% \end{figure}

%\sout{The experiments are performed on a Fedora 27 machine with Intel Dual Core i7@2.0 GHz CPU, 8 GB of RAM, and Python 3.6.}

\subsection{Experimental setup for dispatcher evaluation}

\begin{figure}[!t]
  \begin{json}
  {
	"system_name": "Seth - HPC2N",
	"start_time": 1027839845,
	"equivalence": {
		"processor": {
			"core": 2
		}
	},
	"groups": {
		"g0": {
			"core": 4, 
			"mem": 1000000
		}
	},
	"resources": {
		"g0": 120
	}
  }
  \end{json}
  \caption{System configuration of Seth.}
  \label{fig:json-seth}
\end{figure}

To conduct the experimental study regarding  dispatcher evaluation, we use the Seth dataset introduced in Section \ref{sec:comparison}, given its reasonable size for proof of concept. The corresponding  synthetic system configuration is shown in Figure \ref{fig:json-seth}. % and \ref{fig:json-ricc}. 
Since multiple jobs can co-exist on the same node, we consider a better representation of the system, made of cores instead of processors. \zeynep{We note that AccaSim can as well be used to simulate an HPC system possessing heterogeneous resources, such as the Eurora system, as was shown in \cite{DBLP:conf/supercomputer/NettiGKSB18}.}

As for dispatchers, we employ all the implemented and available dispatchers of AccaSim which are composed of all combinations between the schedulers: First In First Out (FIFO), Shortest Job First (SJF), Longest Job First (LJF) and Easy Backfilling with FIFO priority (EBF); and the allocators:  First Fit (FF) and Best Fit (BF). 
To run the experiments, we conveniently use the \emph{experimentation} tool of AccaSim, as was shown in Figure \ref{fig:implementation-exp}. Each experiment corresponds to the simulation of the Seth workload using a specific dispatcher, and is repeated 10 times so as to obtain reliable and representative results. 

\subsection{Dispatcher evaluation}

\begin{figure}[!t]
\centering
\begin{minipage}{.5\textwidth}
  \centering
    \includegraphics[width=1\textwidth,valign=t]{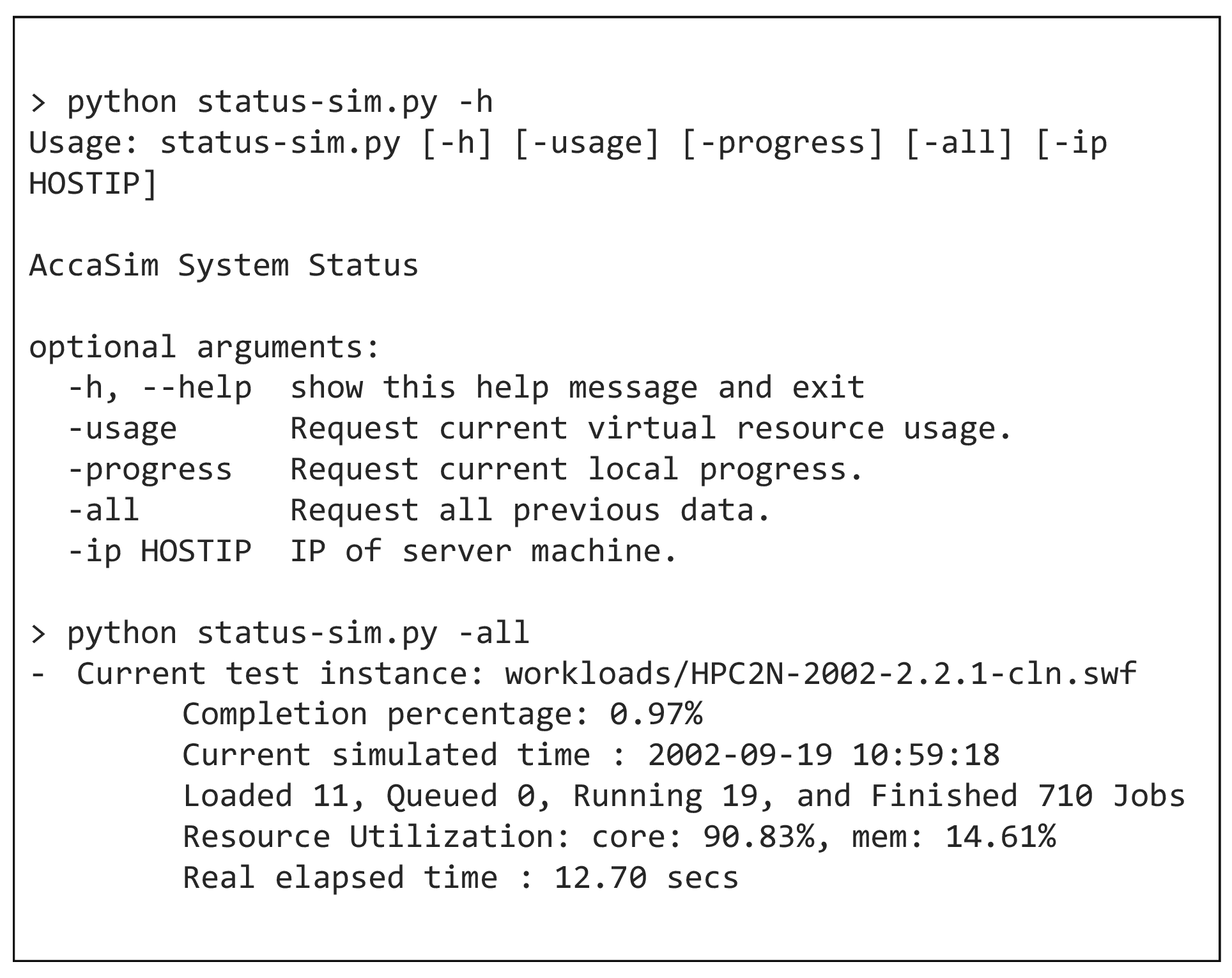}
    \caption{System status.}
    \label{casestudy:watcher}
\end{minipage}%\quad
\begin{minipage}{.5\textwidth}
  \centering
     \includegraphics[width=1\textwidth,valign=t]{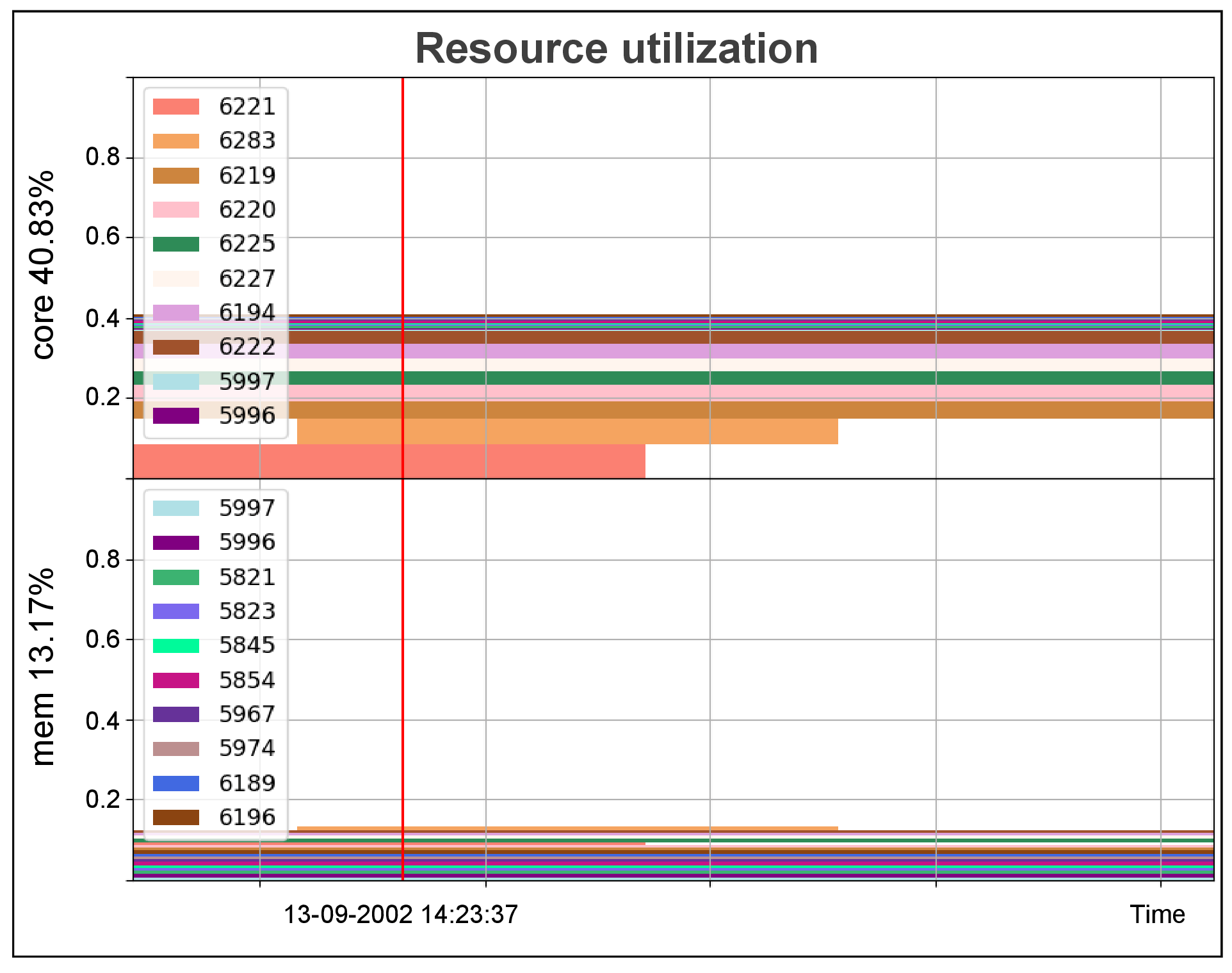}
     \caption{System visualization.}
     \label{casestudy:visualization}
\end{minipage}
\end{figure}

% \begin{figure}[!t]
% 	\centering
%     \includegraphics[width=0.5\textwidth,trim=129 169 128 170,clip=true,angle=90,valign=t]{../figures/Fig9.eps}
%     \caption{System status.}
%     \label{casestudy:watcher}
% \end{figure}

Dispatchers can be evaluated and compared from different perspectives thanks to AccaSim's tools and output data. In Figures \ref{casestudy:watcher} and \ref{casestudy:visualization}, sample snapshots taken by the two components of the \emph{monitoring} tool at certain time points during the FIFO-FF experiment are shown. The \emph{system status} tool receives command line queries to show a variety of information regarding the current synthetic system status, such as the queued jobs, the running jobs, the completed jobs, resource utilization, the current simulation time point, as well as the total CPU time elapsed by the simulator. The \emph{system visualization} tool summarizes the allocation of resources by the running jobs each indicated with a different color, using an estimation (such as wall-time) for job duration. The display is divided by the types of synthetic resources. In our case study, the core and memory usage are shown separately. 

% \begin{figure}[!t]
% 	\centering
%     \includegraphics[height=.48\textwidth,trim=127 167 126 168,clip=true,angle=90,valign=t]{../figures/Fig10.eps}
%     \caption{System visualization.}
%     \label{casestudy:visualization}
% \end{figure}

%While AccaSim users can analyze the output data as they wish, 
The \emph{experimentation} tool automatically generates plots to compare the dispatchers according to their effect on system utilization, job response times, system throughput, and their performance in terms of the time they incur for generating a decision. For job response times and system throughput, two metrics are used. The first is the job slowdown, a common indicator for evaluating job scheduling algorithms \cite{feitelson2001metrics}, which quantifies the effect of a dispatching method on the jobs themselves and is directly perceived also by the HPC users. The slowdown of a  job $j$ is a normalized response time and is defined as $slowdown_j = (T_{w,j} + T_{r,j})/{T_{r,j}}$ where $T_{w,j}$ is the waiting time and $T_{r,j}$ is the duration of job $j$. A job waiting more than its duration  has a higher slowdown than a job waiting less than its duration.  The second metric is the queue size, which counts the number of queued jobs at a certain dispatching time. This metric is a measure of the effects of dispatching on the computing system itself. The lower these two metrics are, the better job response times and system throughput are. 

% \begin{figure}[!t]
% 	\centering
%     \includegraphics[width=0.48\textwidth, trim={0 5 0 0},clip=true]{../figures/Fig11.eps}
% 	  \caption{Distributions for job slowdown.}
%     \label{casestudy:slowdowns}
% \end{figure}

% \begin{figure}[!t]
% 	\centering
%     \includegraphics[width=.48\textwidth, trim={0 5 0 0},clip=true]{../figures/Fig12.eps}
% 	  \caption{Distributions of queue size.}
%     \label{casestudy:queuesize}
% \end{figure}

\begin{figure}[!t]
\captionsetup{justification=raggedright}  
\centering
\begin{minipage}[t]{.5\textwidth}
  \centering
     \includegraphics[width=1\textwidth,valign=t]{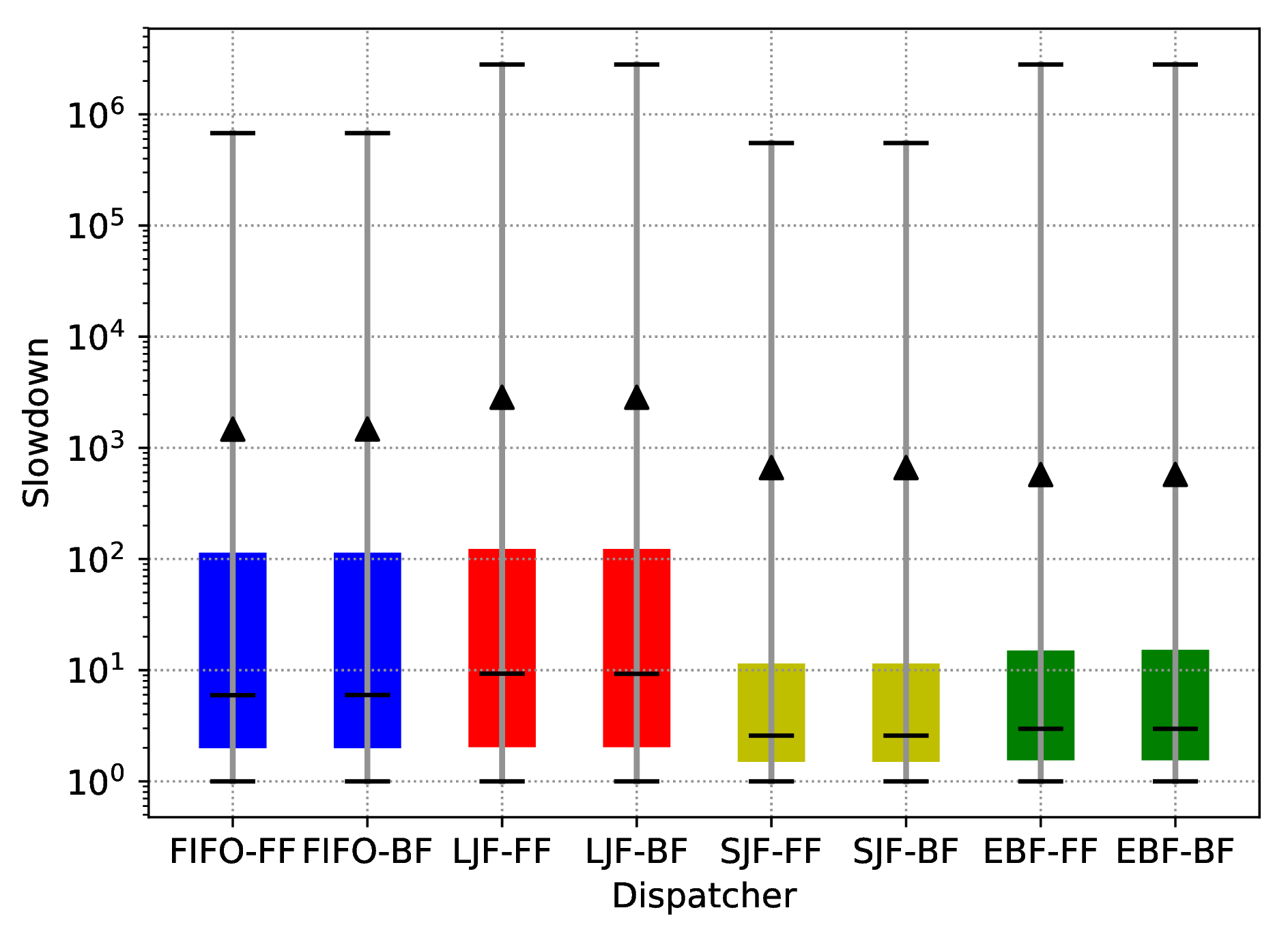}
	 %\caption{Distributions for job slowdown.}
     %\label{casestudy:slowdowns}
\end{minipage}%\quad
\begin{minipage}[t]{.5\textwidth}
  \centering
     \includegraphics[width=1\textwidth,valign=t]{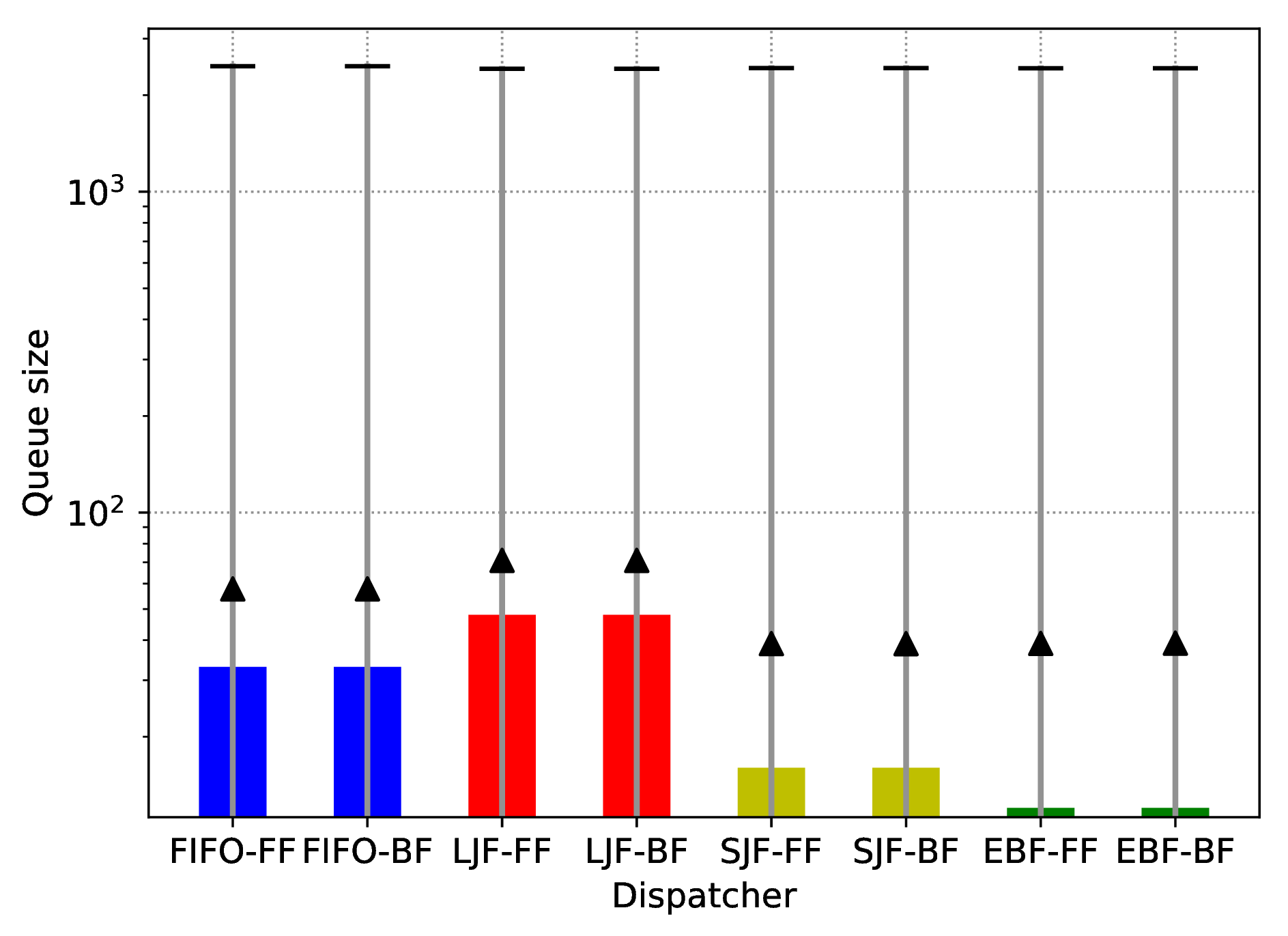}
 	 % \caption{Distributions of queue size.}
     %\label{casestudy:queuesize}
\end{minipage}
\par
\begin{minipage}[t]{.5\textwidth}
  \centering
	 \caption{Distributions for job slowdown.}
     \label{casestudy:slowdowns}
\end{minipage}%\quad
\begin{minipage}[t]{.5\textwidth}
  \centering
 	  \caption{Distributions of queue size.}
     \label{casestudy:queuesize}
\end{minipage}
\end{figure}

In Figures \ref{casestudy:slowdowns} and \ref{casestudy:queuesize}, we present the automatically-generated box-and-whisker plots showing the distributions of the slowdown and the queue size for each experiment. We can see that SJF and EBF-based dispatchers achieve the best results, independently of their allocators probably due to the homogeneous nature of the synthetic system. Their slowdown values are mainly lower than the median of the FIFO and LJF-based dispatchers. SJF maintains overall lower slowdown values than the others, but a higher mean than the EBF. SJF maintains also slightly higher mean in the queue size than the EBF. The scheduling algorithm of EBF does not sort the jobs, like SJF, instead it tries to fit as many jobs as possible into the system, which can explain the best average results achieved in terms of slowdown and queue size.

\begin{figure}[!t]
\captionsetup{justification=raggedright}  
\centering
\begin{minipage}[t]{.5\textwidth}
  \centering
    \includegraphics[width=1\textwidth,valign=t]{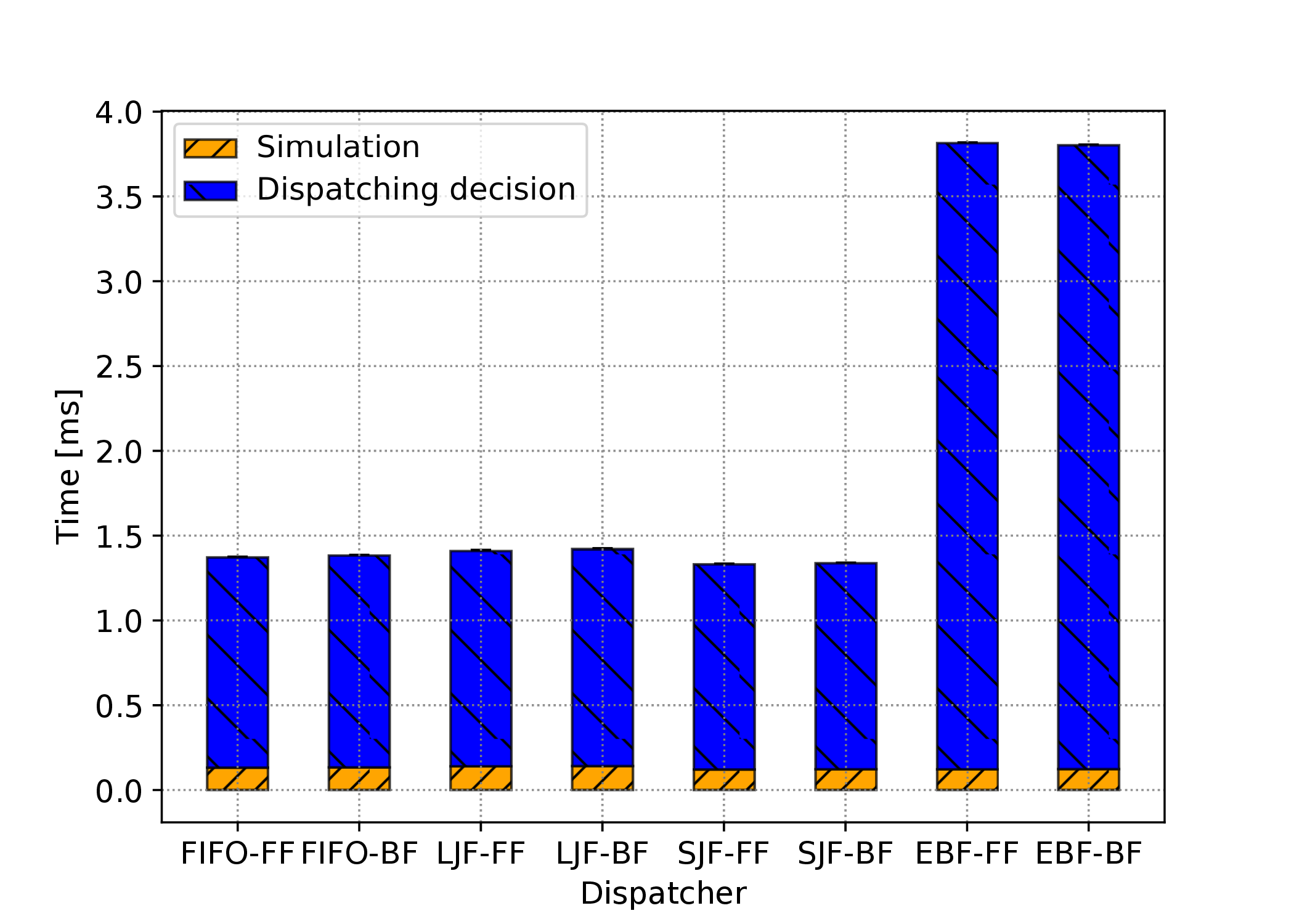}
 	%\caption{Average CPU time at a simulation time point.}
 	%\label{casestudy:timeperstep}
\end{minipage}%\quad
\begin{minipage}[t]{.5\textwidth}
  \centering
     \includegraphics[width=1\textwidth,valign=t]{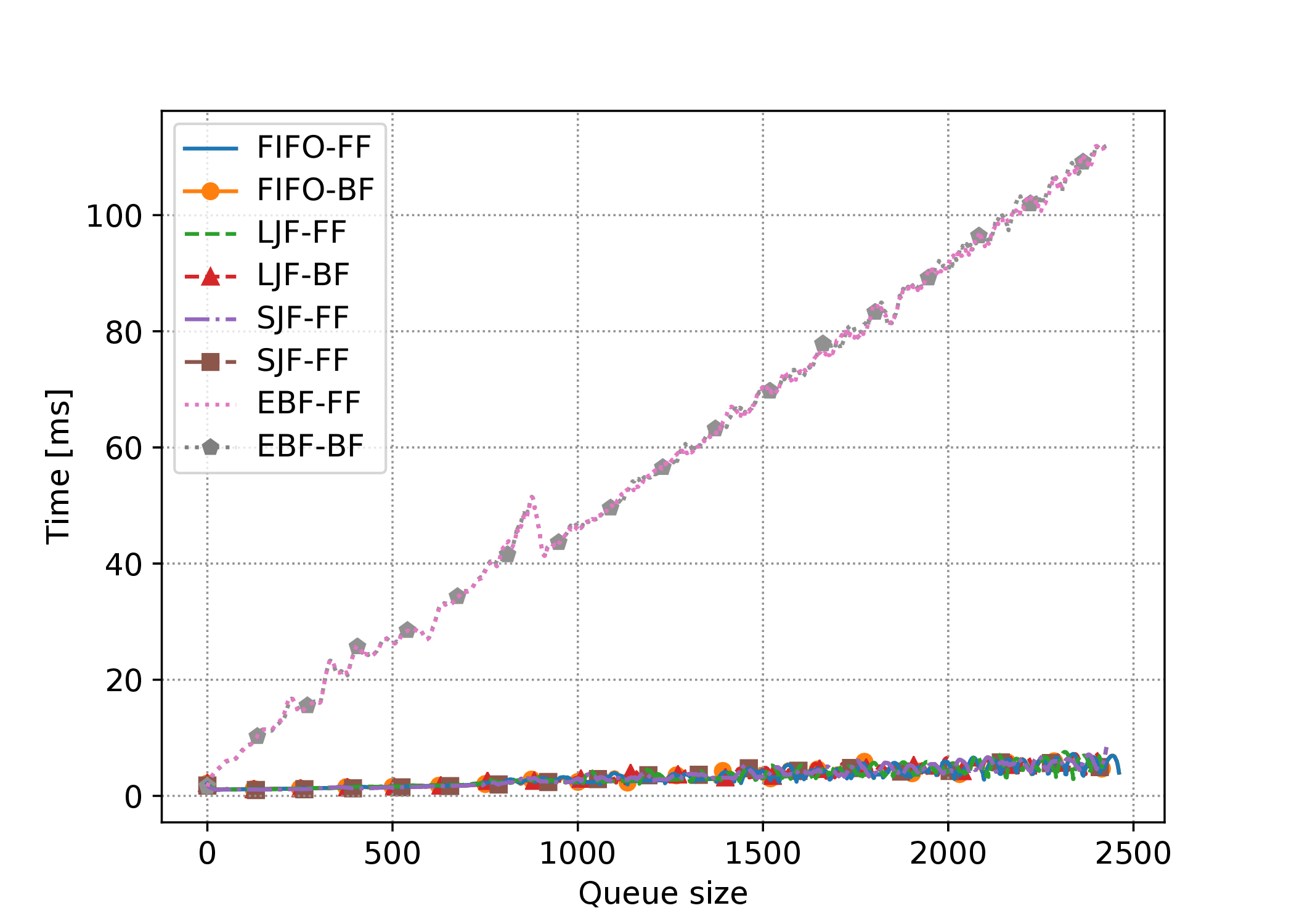}
     %\caption{Average CPU time at a simulation time point to generate a dispatching decision w.r.t. queue size.}
     %\label{casestudy:timepersize}
\end{minipage}
%\par
\begin{minipage}[t]{.5\textwidth}
 	\caption{Average CPU time at a simulation time point.}
 	\label{casestudy:timeperstep}
\end{minipage}%\quad
\begin{minipage}[t]{.5\textwidth}
     \caption{Average CPU time at a simulation time point to generate a dispatching decision w.r.t. queue size.}
     \label{casestudy:timepersize}
\end{minipage}
\end{figure}

In Figure \ref{casestudy:timeperstep}, we present the automatically-generated plot which shows the average CPU time required at a simulation time point for each dispatcher, after aggregating data from all 10 iterations of the experiments. %\sout{In accordance with  Table \ref{casestudy:usagestats},} 
The time spent in simulation, other than generating the dispatching decision, is constant (around 0.2 ms) across all the experiments, and the EBF-based dispatchers spend much more time in generating a decision than the others. In Figure \ref{casestudy:timepersize}, we instead present the automatically-generated plot that analyzes the scalability. Specifically, it reports for each queue size the average CPU  time spent at a simulation time point in generating a dispatching decision. Also in this case, we considered the data related to all 10 iterations of the experiments. While all the dispatchers scale well, the EBF-based dispatchers require more CPU time for processing bigger queue sizes, due to their scheduling algorithm which tries to fit as many jobs as possible into the system. 

% \begin{figure}[!t]
%   	\centering
% 	\includegraphics[width=.48\textwidth,trim={0 4 50 39},clip=true]{../figures/Fig13.eps}
% 	\caption{Average CPU time at a simulation time point.}
% 	\label{casestudy:timeperstep}
% \end{figure}

% \begin{figure}[!t]
% 	\centering
%     \includegraphics[width=.48\textwidth,trim={0 7 48 39}, clip=true]{../figures/Fig14.eps}
%     \caption{Average CPU time at a simulation time point to generate a dispatching decision w.r.t. queue size.}
%     \label{casestudy:timepersize}
% \end{figure}

\begin{table*}[!ht]
\centering
\begin{tabular}[t]%{|c|C{0.6cm}|C{0.2cm}|C{0.6cm}|C{0.2cm}|C{0.6cm}|C{0.2cm}|C{0.6cm}|C{0.2cm}|}
{|c|c|c||c|c||c|c||c|c|}
\hline
\multirow{3}{*}{Dispatcher} & \multicolumn{4}{c||}{Time (MM:SS)} & \multicolumn{4}{c|}{Memory (MB)} \\ \cline{2-9}
& \multicolumn{2}{c||}{Total} & \multicolumn{2}{c||}{Disp.} & \multicolumn{2}{c||}{Avg.} & \multicolumn{2}{c|}{Max.}  \\ \cline{2-9}
& \(\mu\) & \(\sigma\) & \(\mu\) & \(\sigma\) & \(\mu\) & \(\sigma\) & \(\mu\) & \(\sigma\)  \\ \hline
FIFO-FF & 08:01 & 2.6 & 07:15 & 2.3 & 76 & 0.2 & 82 & 0.3 \\ \hline
FIFO-BF & 08:05 & 1.8 & 07:18 & 1.6 & 79 & 0.1 & 85 & 1.1 \\ \hline
LJF-FF & 08:13 & 2.4 & 07:24 & 2.1 & 80 & 0.7 & 86 & 0.9 \\ \hline
LJF-BF & 08:17 & 2.3 & 07:27 & 2.1 & 81 & 0.8 & 86 & 0.9 \\ \hline
SJF-FF & 07:46 & 2.2 & 07:04 & 2.0 & 82 & 0.8 & 86 & 0.5 \\ \hline
SJF-BF & 07:49 & 1.7 & 07:06 & 1.5 & 82 & 0.4 & 86 & 0.6 \\ \hline
EBF-FF & 22:24 & 2.9 & 21:41 & 2.7 & 82 & 0.6 & 85 & 0.7 \\ \hline
EBF-BF & 22:19 & 4.6 & 21:36 & 4.2 & 82 & 0.6 & 84 & 0.8 \\ \hline
 \end{tabular}	
 \caption{Total CPU time and memory usage during the simulation.}
 \label{casestudy:usagestats}
\end{table*}

AccaSim users are free to analyze the output data as they wish to evaluate the dispatchers further. For instance, to compare in more detail the dispatchers' performance, they can extract the total usage of CPU time and memory of each experiment, as reported in Table~\ref{casestudy:usagestats}. In the table, the time columns correspond to the total CPU time spent by the simulator and the time spent in generating the dispatching decision; whereas the memory columns give the average and the maximum amount of memory utilized over the entire simulation time points. The reported values of an experiment are aggregated across all the 10 iterations, and both mean (\(\mu\)) and standard deviation (\(\sigma\)) are shown.
%\sout{In the first thread, FIFO and LJF-based experiments are completed in around 59 minutes, while in the second, the SJF and EBF-based experiments are completed in around 70 minutes.}

Most of the experiments took around 8 minutes. The exceptions are the EBF-based experiments which require around 22 minutes because the underlying dispatching algorithms are computationally more intensive. In accordance with Figure  \ref{casestudy:timeperstep}, the time spent by the simulator, other than generating the dispatching decision, is constant (around 40 seconds) across all the experiments. The total CPU usage is thus highly dependent on the complexity of the dispatcher. %As for the memory usage, thanks to the incremental job loading and job removal  capabilities,  AccaSim consumes low memory.  
The average memory usage is around 80MB with a peak at 86MB across all the experiments. 
%\zeynep{Such low memory usage makes it possible to execute experiments in parallel. Considering the  size of the dataset, these numbers are very reasonable, supporting the claim that AccaSim is scalable.}

Our analysis restricted to the considered dataset reveals that,  while the EBF-based dispatchers give the best results in terms of response times and throughput, they are much more costly in generating a dispatching decision. Simple dispatchers based on SJF are valid alternatives with their excellent scalability and with their comparable results in response times and throughput.

\subsection{Synthetic workload datasets}

In order to generate synthetic workload datasets, and later for comparison purposes, we utilize the Seth and RICC datasets introduced in Section \ref{sec:comparison}. With each, we generate four datasets using different configurations in terms of resource type, processing unit performance, and the number of jobs. The first dataset includes 50,000 jobs and a 1.5x improvement in core performance. The second includes 100,000 jobs with double number of nodes. The third includes 200,000 jobs, two GPU accelerator cards for a quarter of the nodes with a performance of 933 GFLOPS per second. Finally, the last includes 500,000 jobs, two GPU accelerator cards for a half of the nodes with a performance of 933 GFLOPS per second and a 1.5x improvement in the core performance. The improved performance and the change in the number of nodes are relative to the system  that the workload dataset in consideration belongs to. 
In the following, we first briefly describe the generation process, and then show the similarity between the real and the generated datasets. 

\paragraph{Synthetic workload dataset generation.}

The first aspect to compare between a real and a synthetic workload dataset is the job submission cycle which refers to the job submission times and reflects the usage of the system by its users. The cycles could be represented by  certain periods of working time to reflect better the real usage of the system. 
The \emph{WorkloadGenerator} calculates the submission time of a job $j$ based on a daily cycle model proposed in \cite{DBLP:journals/jpdc/LublinF03}. 
In the original algorithm, named Slot Weight Method, a day is represented by 48 slots of 30 minutes each ($s$). Thus, the first slot starts at midnight, the next one at 00:30, and so on. Each slot has a specific weight which is the ratio between the number of jobs  belonging to the time slot and the total number of jobs in the real workload dataset, which represents a measure for selecting a slot for $j$.
The algorithm generates a random value $v$ between 0 and 5 to represent the maximum number of days that can elapse between $j$ and its predecessor, based on the statistical distribution of the interarrival times of the real workload dataset.
For selecting a slot, the algorithm starts from the slot of the predecessor of $j$. The slots are considered as a circular list. For each considered slot, if $v$ is greater or equal to the slot weight, $v$ is updated by subtracting the slot weight. Update continues with the next slot, otherwise, the algorithm stops and selects the current slot. Then, the job submission time of $j$ is calculated by summing the half hours of all the surpassed slots plus the remaining amount of $v$. 

We modify this algorithm in two aspects so as to assimilate a real job submission cycle. First, we modify the fixed upper-bound $v_{max}$ of $v$ to the maximum value of the interarrival times of the dataset. Second, we add a dynamic process that modifies $v_{max}$ during a job submission time generation.  For this purpose, we calculate the ratio between the number of the currently generated jobs and the required jobs in three different ways in relation to the last submitted hour, the last submitted day, and the last submitted month. This allows to keep the generation of values as similar to the real data as possible. Then, we calculate the progress ratio of each ratio by dividing it by the respective ratios in the real data. The overall progress ratio is the multiplication between all progress ratios ($pr$). Finally, $v_{max}$ is dynamically adapted at each job submission time generation as follows: 
 $$
 v_{max} \leftarrow v_{max} - (v_{max} - s) * (1 - pr)
 $$
If $pr = 1$, the  job submission time generation of the  predecessor reached the real ratios, thus for $j$, we use $v_{max}$. In addition, when the real data does not include specific months, $pr$ has only hourly and daily ratios.

The second aspect to compare is the theoretical computed FLOPs for each job during its execution in the system, which depends on, among others, its duration and resource requests in terms of resource type  restricted to the processing units (e.g., cores, GPU, MIC, etc.) and quantity.
These features of a job are generated in three phases. The first phase is based on an algorithm from \cite{DBLP:journals/jpdc/LublinF03} to select the job type, serial or parallel, and the number of requested nodes. Since this algorithm considers a job parallel if it runs on multiple nodes, we modify it to create parallel jobs on a single node, i.e. when the number of required cores is greater than one. In the second phase, the resource request  is defined by randomly choosing among the available resource types and assigning them a quantity, using a uniform distribution and considering the request limits passed as an argument during the \emph{WorkloadGenerator} instantiation, as shown in Figure \ref{fig:wgen_instantiation}. Finally, in the third phase, the job duration is calculated as the division between (i) a random FLOP value and (ii) the dot product of the resource requests and their corresponding theoretical performance, multiplied by the number of required nodes. 

% \begin{figure}[!t]
%  \centering
%  \includegraphics[width=.48\textwidth,trim=30 27 50 23,clip=true]{figures/Fig15.eps}
% 	\caption{Seth workload dataset.}
% 	\label{workload:sub-hp2cn}
% \end{figure}

% \begin{figure}[!t]
% \centering
% 	\includegraphics[width=.48\textwidth,trim=30 27 50 23,clip=true]{figures/Fig16.eps}
% 	\caption{RICC workload dataset.}
% 	\label{workload:sub-ricc}
% \end{figure}

\begin{figure}[!t]
\centering
\begin{minipage}{.5\textwidth}
  \centering
  \includegraphics[width=1\textwidth,valign=t]{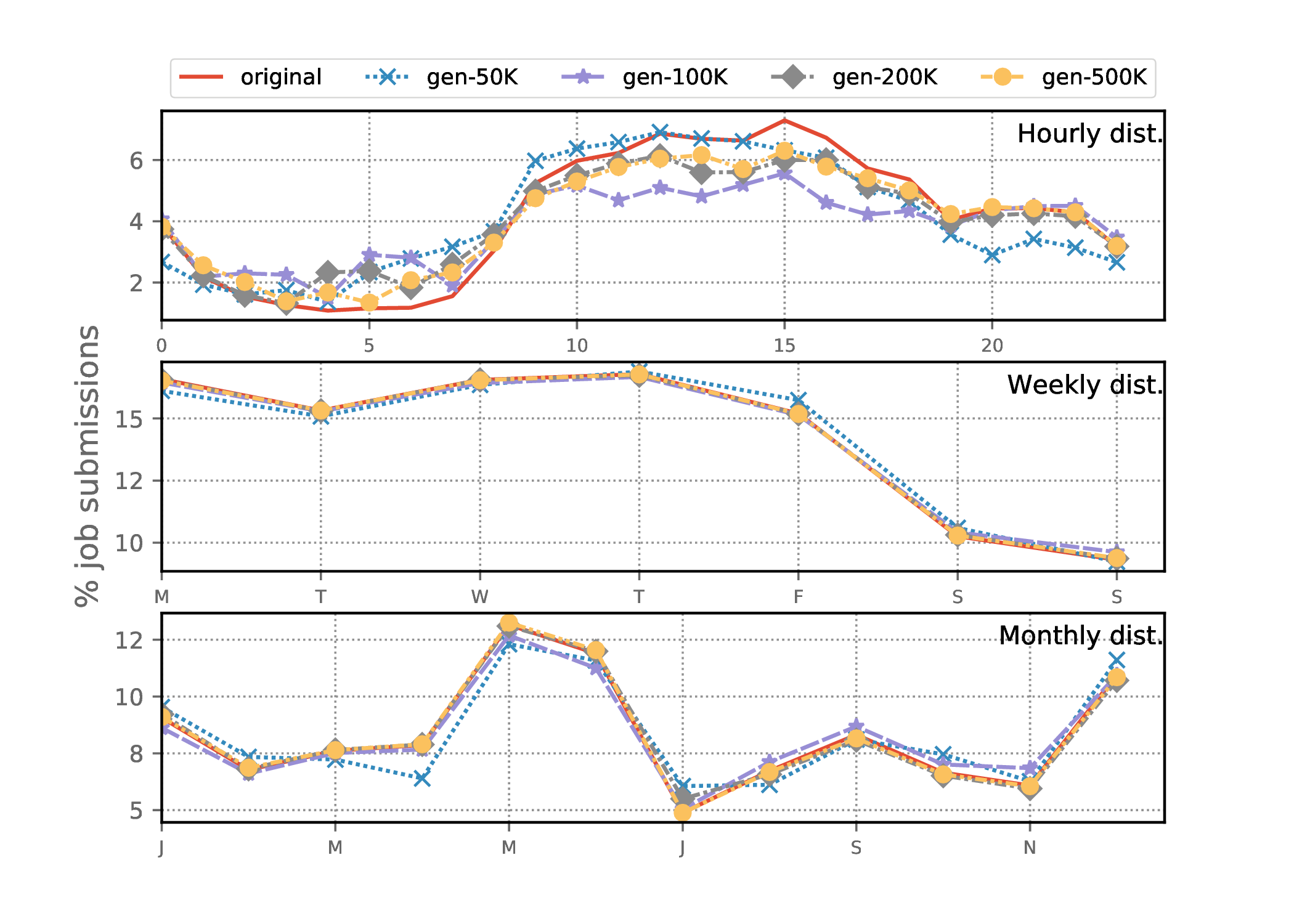}
 	\caption{Seth workload dataset.}
 	\label{workload:sub-hp2cn}
\end{minipage}%\quad
\begin{minipage}{.5\textwidth}
  \centering
    \includegraphics[width=1\textwidth,valign=t]{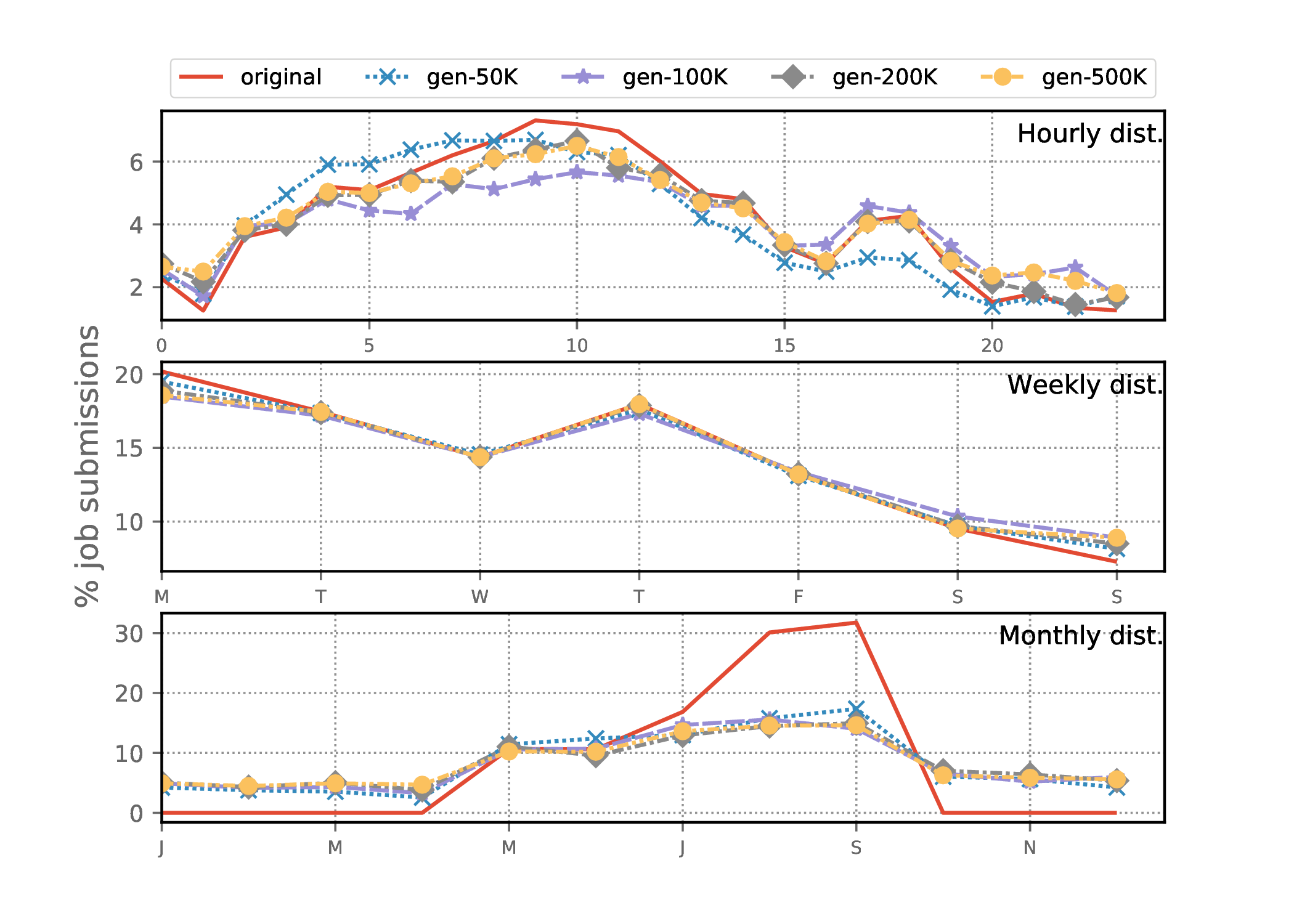}
 	\caption{RICC workload dataset.}
 	\label{workload:sub-ricc}
\end{minipage}
\end{figure}

\paragraph{Comparison to the real workload datasets.}

Figures \ref{workload:sub-hp2cn} and \ref{workload:sub-ricc} show the the hourly, daily, monthly  job submission distributions of  the real and the generated workload datasets. The introduced modifications generate submissions that took place mainly during the working hours, weekdays, and working months, resulting in a more realistic scenario. The generated datasets look very similar to the real datasets, except in the case of the monthly distribution of the RICC dataset. The reason is that the RICC job submissions span to five months, not to an entire year. 

% \begin{figure}[!t]
% 	\centering
% 	\includegraphics[width=.48\textwidth,trim=17 7 50 43,clip=true]{figures/Fig17.eps}
% 	\caption{Seth workload dataset.}
% 	\label{workload:flop-hp2cn}
% \end{figure}

% \begin{figure}[!t]
% 	\centering
% 	\includegraphics[width=.48\textwidth,trim=17 7 50 43,clip=true]{figures/Fig18.eps}
% 	\caption{RICC workload dataset.}
% 	\label{workload:flop-ricc}
% \end{figure}

\begin{figure}[!t]
\centering
\begin{minipage}{.5\textwidth}
  \centering
 	\includegraphics[width=1\textwidth]{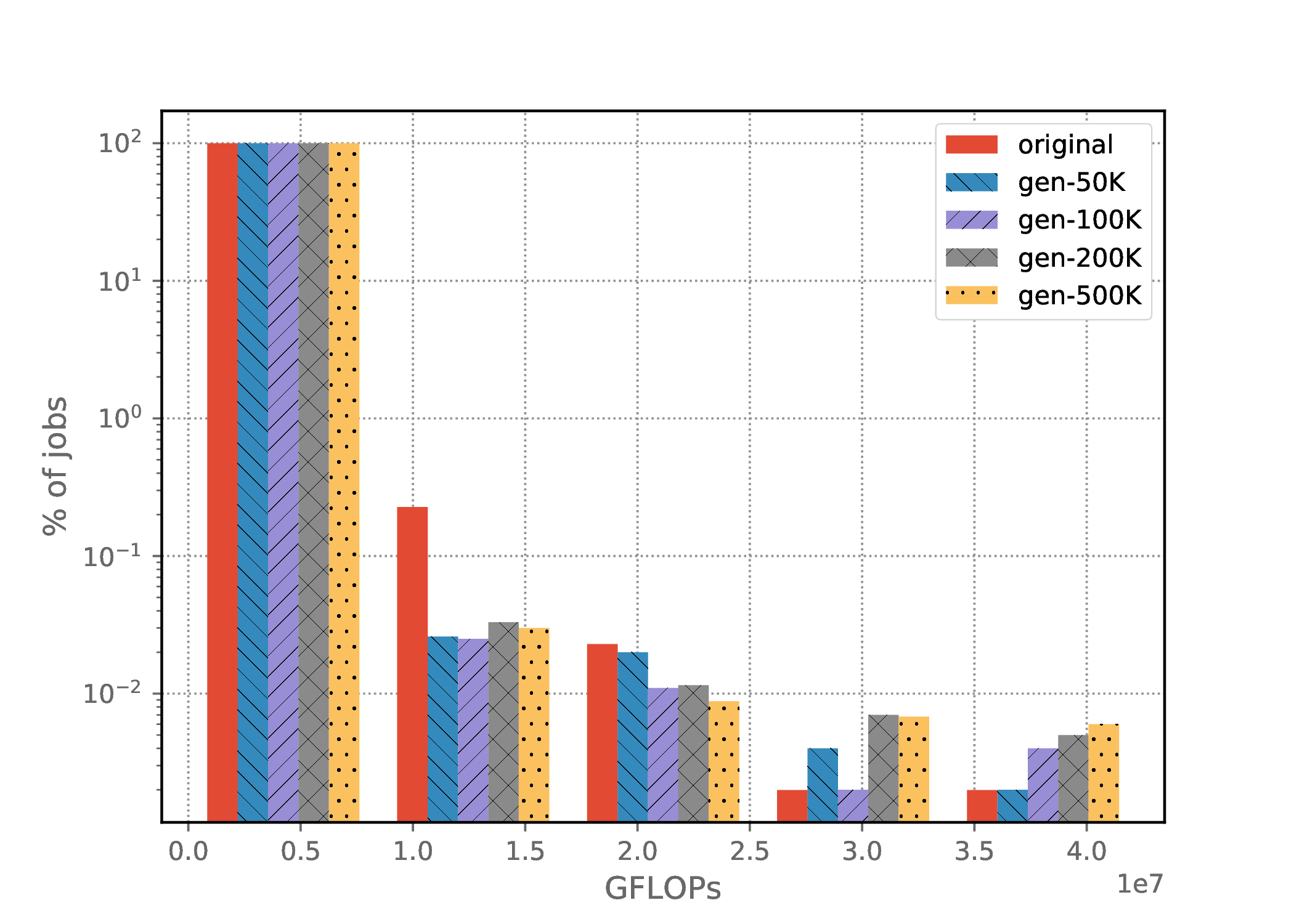}
 	\caption{Seth workload dataset.}
 	\label{workload:flop-hp2cn}
\end{minipage}%\quad
\begin{minipage}{.5\textwidth}
  \centering
    \includegraphics[width=1\textwidth,valign=t]{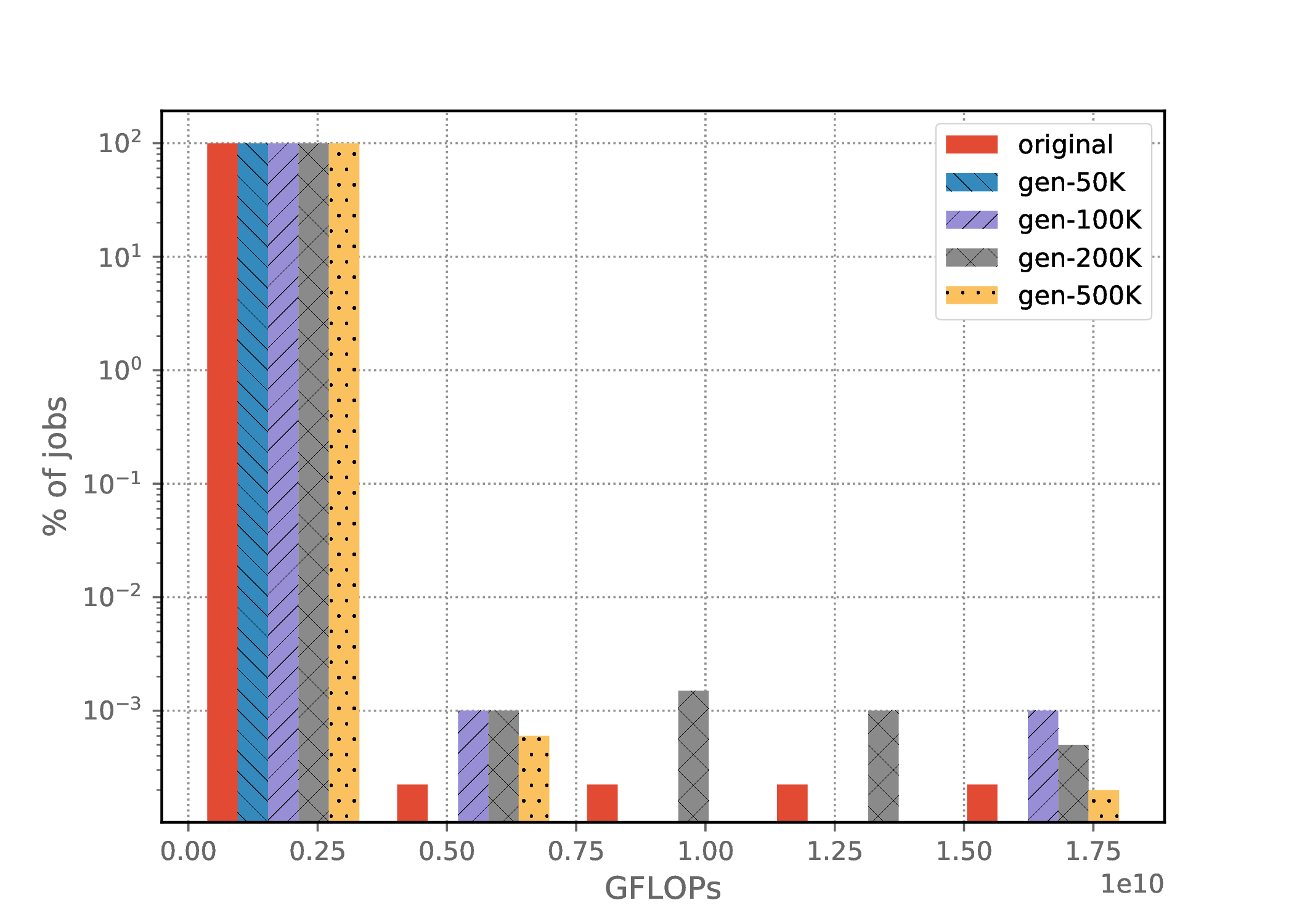}
 	\caption{RICC workload dataset.}
 	\label{workload:flop-ricc}
\end{minipage}
\end{figure}

Figures \ref{workload:flop-hp2cn} and \ref{workload:flop-ricc} show the distributions of the  computed theoretical FLOPS, here represented in GFLOPS, between the real and the generated workload datasets. We observe a similar pattern also here. 
The usage of the FLOPS calculation for the generation of the jobs' features allows maintaining a distribution similar to the real workload dataset, independent of the configuration of the real system. In this way, the real dataset can be tested with other system configurations using the generated dataset. 

%% file: sections/conclusions.tex
\section{Conclusions}\label{sec:conclusion}

In this paper, we presented AccaSim, a library for simulating WMS in an HPC system, which offers to the researchers an accessible tool to facilitate their job dispatching research.  The library is open-source, implemented in Python, which is freely available for any major operating system, and works with dependencies reachable in any distribution. It is executable on a wide range of computers thanks to its lightweight installation and light memory footprint. AccaSim is scalable  to large workload datasets and provides support for easy customization, allowing to carry out experiments across different workload sources, resource types, and dispatching algorithms. Moreover, AccaSim enables users to develop novel advanced dispatchers by exploiting  information regarding the current system status, which can be extended for including custom behaviors such as energy and power consumption and failures of the resources. Last but not least, AccaSim aids users in their experiments via automated tools  to generate synthetic workload datasets, to run the simulation experiments and to produce plots to evaluate dispatchers.  The researchers can thus use AccaSim to mimic any real system, \zeynep{including those possessing heterogeneous resources}, develop advanced dispatchers using for instance power and energy-aware, fault-resilient algorithms, and test and  \zeynep{evaluate} them in a convenient way over a wide range of  workload sources by using real workload traces or by generating them. 

\zeynep{In order to highlight the main contributions of AccaSim, we discussed the existing related simulators, presented  a critical comparison to the most similar simulators, and showcased AccaSim's use in job dispatching research, specifically in dispatcher evaluation and synthetic workload generation. 
%We believe that AccaSim is an attractive tool for conducting job dispatching research in HPC systems.
In future work, we plan to use AccaSim to develop advanced dispatchers using power and energy-aware, fault-resilient algorithms. }